\documentclass[twocolumn,english]{article}
\usepackage{ae,aecompl}
\usepackage[T1]{fontenc}
\usepackage[latin9]{inputenc}
\usepackage[a4paper]{geometry}
\usepackage{color}
\usepackage{babel}
\usepackage{bm}
\usepackage{amsmath}
\usepackage{amssymb}
\usepackage{graphicx}
\usepackage[unicode=true]
 {hyperref}

\makeatletter
\newcommand{\lyxaddress}[1]{
	\par {\raggedright #1
	\vspace{1.4em}
	\noindent\par}
}

\usepackage{xcolor}            
\definecolor{myred}{rgb}{0.66, 0.15, 0.15}

\usepackage{hyperref}          
\hypersetup{
     colorlinks   = true,
     citecolor    = myred,
     urlcolor     = myred,
     urlbordercolor = myred
}

\usepackage[normalem]{ulem} 
\usepackage{flushend}
\usepackage{cuted}

\makeatother

\begin{document}
\begin{strip}

\textsf{\textbf{\Huge{}Width and shift of Fano-Feshbach}}{\Huge\par}

\textsf{\textbf{\Huge{}resonances for van der Waals interactions}}{\Huge\par}

{\Large{}\vspace{0.8cm}
Pascal Naidon$^{1}$ and Ludovic Pricoupenko$^{2}$}{\Large\par}

{\Large{}\vspace{0.5cm}
}\today\\
\vspace{0.5cm}

\lyxaddress{{\small{}$^{1}$RIKEN Nishina Centre, Quantum Hadron Physics Laboratory,
RIKEN, Wak{\=o}, 351-0198 Japan.}\\
{\small{}}\textit{\href{mailto:pascal@riken.jp}{pascal@riken.jp}}}

\lyxaddress{{\small{}$^{2}$Sorbonne Université, CNRS, Laboratoire de Physique
Théorique de la Matière Condensée (LPTMC), F-75005 Paris, France.}\\
\emph{\small{}}\textit{\href{mailto:ludovic.pricoupenko@sorbonne-universite.fr}{ludovic.pricoupenko@sorbonne-universite.fr}}}
\begin{abstract}
We revisit the basic properties of Fano-Feshbach resonances in two-body
systems with van der Waals tail interactions, such as ultracold neutral
atoms. Using a two-channel model and two different methods, we investigate
the relationship between the width and shift of the resonances and
their dependence on the low-energy parameters of the system. Unlike
what was previously believed {[}\href{http://link.aps.org/doi/10.1103/RevModPhys.82.1225}{Rev. Mod. Phys. 82, 1225 (2010)}{]}
for magnetic resonances, we find that the ratio between the width
and the shift of a resonance does not depend only on the background
scattering length, but also on a closed-channel scattering length.
We obtain different limits corresponding to different cases of optical
and magnetic resonances. Although the generalisation of the theory
to the multi-channel case remains to be done, we found that our two-channel
predictions are verified for a specific resonance of lithium-6.
\end{abstract}
\end{strip}

\section{Introduction}

\hypersetup{linkcolor=myred}

A Fano-Feshbach resonance~\cite{Feshbach1958,Fano1961} is the strong
modification of the scattering properties of two particles due to
their coupling with a bound state in a different internal state. At
low energy where the s-wave scattering is dominant, these resonances
cause the scattering length of the two particles to diverge. While
such resonances may accidentally occur in nature~\cite{Verhaar1993},
it was realised that they could be induced in ultracold alkali atoms
by applying a magnetic field to these systems~\cite{Tiesinga1993}.
Because of different Zeeman shifts experienced by different hyperfine
states of atoms, it is possible to tune the intensity of the magnetic
field such that a bound state in a certain hyperfine state approaches
the scattering energy of the two atoms, resulting in a Fano-Feshbach
resonance. This led to one of the major achievements in the field
of ultracold atoms, the possibility to control their interactions,
enabling the experimental study of a wealth of fundamental quantum
phenomena for over nearly two decades~\cite{Inouye1998,OHara2002,Regal2003,Bourdel2004,Kraemer2006,Haller2009,Schirotzek2009,Makotyn2014,Hu2016}.

The general formalism of Fano-Feshbach resonances has already been
studied in detail~\cite{Feshbach1958,Fano1961,Joachain1983,Chin2010}.
This work focuses on the general relationship between the width and
shift characterising Fano-Feshbach resonances. In section \ref{sec:2}
of this article, we introduce the two-channel model which is used
afterwards to derive analytic relations. In section \ref{sec:3} we
recall how the shift and width of the resonance can be deduced in
the isolated resonance approximation. In section \ref{sec:4} and
\ref{sec:5}, we establish the relationship between the shift $\Delta$
and the width $\Gamma$, in particular for systems characterised at
large interparticle distance by a van der Waals interaction. 

Our main result shows the dependence of these quantities upon the
open-channel (background) scattering length $a_{{\rm bg}}$ and a
closed-channel scattering length $a_{{\rm c}}$:

\begin{align}
\Delta & \propto\left(a_{{\rm bg}}-a_{{\rm c}}\right),\qquad\Gamma\propto\left(a_{{\rm bg}}-a_{{\rm c}}\right)^{2}.\label{eq:MainResult}
\end{align}
This result is inconsistent with the formula given by Eq.~(37) of
Ref.~\cite{Chin2010} obtained from multi-channel quantum defect
theory (MQDT). To clarify this discrepancy, in section \ref{sec:6}
we use the MQDT approach to rederive the width and shift. This derivation
turns out to confirm our results obtained with the isolated resonance
approximation. Moreover, we show that the formula of Ref.~\cite{Chin2010}
relies on a simplifying assumption that appears to be invalid in general.
Finally, in section \ref{sec:7}, we illustrate our results with the
broad magnetic resonance of lithium-6 atoms.

\section{The two-channel model}

\label{sec:2} The simplest description of Fano-Feshbach resonances
requires two channels, corresponding to two different internal states
of a pair of atoms. Each channel is associated with a different interaction
potential between the two atoms. At large distances, these two potentials
tend to different energies, or thresholds, which are equal to the
energies of two separated atoms in the internal states of the corresponding
channel. For a resonance to occur, the initially separated atoms must
scatter with a relative kinetic energy that is above the threshold
of one channel, called the \textit{open} channel, but below the threshold
of the other channel, called the \textit{closed} channel. In addition,
the relative motion of the atoms in one channel must be coupled to
that of the other channel. The wave function for the relative vector
${\bm{R}}$ between the two atoms with relative kinetic energy $E$
is therefore described by two components ${\Psi_{{\rm o}}(\bm{R})}$
and ${\Psi_{{\rm c}}(\bm{R})}$, respectively for the open and the
closed channel, satisfying the coupled Schr{ö}dinger equations (in
ket notation): 
\begin{align}
 & \left(T+V_{{\rm oo}}-E\right)\vert\Psi_{{\rm o}}\rangle+V_{{\rm oc}}\vert\Psi_{{\rm c}}\rangle=0\label{eq:CC1}\\
 & \left(T+V_{{\rm cc}}-E\right)\vert\Psi_{{\rm c}}\rangle+V_{{\rm co}}\vert\Psi_{{\rm o}}\rangle=0,\label{eq:CC2}
\end{align}
where ${V_{{\rm oo}}}$ and ${V_{{\rm cc}}}$ are the open- and closed-channel
potentials with ${V_{{\rm cc}}(\infty)>E>V_{{\rm oo}}(\infty)}$,
and ${V_{{\rm oc}}=V_{{\rm co}}^{*}}$ are the coupling potentials.
In Eqs.~(\ref{eq:CC1},\ref{eq:CC2}) $T$ is the relative kinetic
energy operator, 
\begin{equation}
T=-\frac{\hbar^{2}}{2\mu}\nabla_{\bm{R}}^{2},\label{eq:KineticEnergyOperator}
\end{equation}
where $\mu$ is the reduced mass of the atoms. For convenience, we
choose ${V_{{\rm oo}}(\infty)=0}$. Equations (\ref{eq:CC1}) and
(\ref{eq:CC2}) can be integrated as follows:
\begin{align}
\vert\Psi_{{\rm o}}\rangle & =\vert\bar{\Psi}_{{\rm o}}^{E}\rangle+G_{{\rm o}}^{+}V_{{\rm oc}}\vert\Psi_{{\rm c}}\rangle\label{eq:IntegratedCC1}\\
\vert\Psi_{{\rm c}}\rangle & =0+G_{{\rm c}}V_{{\rm co}}\vert\Psi_{{\rm o}}\rangle,\label{eq:IntegratedCC2}
\end{align}
where ${G_{{\rm o}}^{+}=\left(E+i0^{+}-T-V_{{\rm oo}}\right)^{-1}}$
and ${G_{{\rm c}}=\left(E-T-V_{{\rm cc}}\right)^{-1}}$ are the resolvents
of the open and closed channels, and ${\vert\bar{\Psi}_{{\rm o}}^{E}\rangle}$
is the scattering eigenstate of the open-channel Hamiltonian ${T+V_{{\rm oo}}}$
at energy $E$. It is energy-normalised, i.e., ${\langle\bar{\Psi}_{{\rm o}}^{E}\vert\bar{\Psi}_{{\rm o}}^{E'}\rangle=\delta(E-E')}$.

\section{Shift and width of an isolated resonance}

\label{sec:3}

The description of a Fano-Feshbach resonance is usually done in the
isolated resonance approximation~\cite{Joachain1983,Timmermans1999,CCT2007}.
In that approximation, only a single bound state ${\vert\Psi_{m}\rangle}$
(here assumed with s-wave symmetry) of the closed channel gives a
significant contribution to the resonance. The closed-channel resolvent
may therefore be decomposed into a resonant and a non-resonant part:
\begin{equation}
G_{{\rm c}}=\frac{\vert\Psi_{m}\rangle\langle\Psi_{m}\vert}{E-E_{m}}+\underbrace{\sum_{n\ne m}\frac{\vert\Psi_{n}\rangle\langle\Psi_{n}\vert}{E-E_{n}}}_{G_{{\rm c}}^{\text{nr}}},\label{eq:GreensFunctionDecomposition}
\end{equation}
where ${\vert\Psi_{n}\rangle}$ and ${E_{n}}$ denote all the eigenstates
and energies of the closed-channel Hamiltonian ${T+V_{{\rm cc}}}$,
normalised as ${\langle\Psi_{n}\vert\Psi_{n^{\prime}}\rangle=\delta_{n,n^{\prime}}}$.
One finds 
\begin{align}
 & \vert\Psi_{{\rm o}}\rangle=\vert\Psi_{\text{{\rm bg}}}\rangle+G_{{\rm o}}^{+}T_{{\rm res}}\vert\Psi_{{\rm bg}}\rangle\label{eq:Psio}\\
 & \vert\Psi_{{\rm c}}\rangle=\vert\Psi_{m}\rangle\frac{\langle\Psi_{m}|V_{{\rm co}}|\Psi_{{\rm o}}\rangle}{E-E_{m}}+G_{{\rm c}}^{{\rm nr}}V_{{\rm co}}\vert\Psi_{{\rm o}}\rangle,\label{eq:Psic}
\end{align}
where we have introduced the background scattering state ${\vert\Psi_{{\rm bg}}\rangle}$
and the operator ${T_{{\rm res}}}$ given by 
\begin{align}
\vert\Psi_{{\rm bg}}\rangle & =\vert\bar{\Psi}_{{\rm o}}^{E}\rangle+G_{{\rm o}}^{+}V_{{\rm oc}}G_{{\rm c}}^{\text{nr}}V_{{\rm co}}\vert\Psi_{{\rm o}}\rangle\label{eq:Psibg}\\
T_{{\rm res}} & =\frac{V_{{\rm oc}}\vert\Psi_{m}\rangle\langle\Psi_{m}\vert V_{{\rm co}}}{E-E_{m}-\langle\Psi_{m}\vert V_{{\rm co}}G_{{\rm o}}^{+}V_{{\rm oc}}\vert\Psi_{m}\rangle}.\label{eq:Tres}
\end{align}
Equation~(\ref{eq:Psio}) shows that ${\vert\Psi_{{\rm o}}\rangle}$
is analogous to a scattering state in a single-channel problem, where
${\vert\Psi_{{\rm bg}}\rangle}$ plays the role of the incident state,
and ${T_{\text{res}}}$ is the transition operator. In this single-channel
picture, the scattering amplitude is thus proportional to the matrix
element ${\langle\Psi_{{\rm bg}}|T_{{\rm res}}|\Psi_{{\rm bg}}\rangle}$
of this transition operator for the incident state. 
From Eqs.~(\ref{eq:Psio}-\ref{eq:Tres}) we have
\begin{equation}
\langle\Psi_{{\rm bg}}|T_{{\rm res}}|\Psi_{{\rm bg}}\rangle=\frac{\Gamma}{2\pi}\frac{1}{E-E_{m}-\Delta+i\Gamma^{\prime}/2},\label{eq:exactAmplitude}
\end{equation}
where $\Delta$ and $\Gamma$ are given by 
\begin{align}
 & \Delta=\langle\Psi_{m}\vert V_{{\rm co}}{\rm Re}{(G_{{\rm o}}^{+})}V_{{\rm oc}}\vert\Psi_{m}\rangle\label{eq:Shift}\\
 & \Gamma=2\pi\vert\langle\Psi_{m}\vert V_{{\rm co}}\vert\Psi_{{\rm bg}}\rangle\vert^{2},\label{eq:Width}
\end{align}
and ${\Gamma^{\prime}=2\pi\vert\langle\Psi_{m}\vert V_{{\rm co}}\vert\bar{\Psi}_{{\rm o}}^{E}\rangle\vert^{2}}$.
In the isolated resonance approximation, whenever the scattering energy
$E$ is close to the molecular energy, the molecular states ${n\ne m}$
only bring a small correction to the closed-channel state in Eq.~\eqref{eq:Psic}
and to the background scattering state in Eq.~\eqref{eq:Psibg}.
One can thus make the approximation ${\vert\Psi_{{\rm c}}\rangle\propto\vert\Psi_{m}\rangle}$
and ${\vert\Psi_{{\rm bg}}\rangle\approx\vert\bar{\Psi}_{{\rm o}}^{E}\rangle}$
yielding ${\Gamma^{\prime}\approx\Gamma}$. We can then identify a
Breit-Wigner law in Eq.~(\ref{eq:exactAmplitude}) with the width
${\Gamma}$ and shift ${\Delta}$. From Eq.~(\ref{eq:Psio}), one
finds the s-wave scattering phase shift, 
\begin{equation}
\eta=\eta_{{\rm bg}}+\eta_{\text{res}}\label{eq:ScatteringPhaseShift}
\end{equation}
where ${\eta_{{\rm bg}}}$ is the background scattering phase shift
contained in ${\vert\psi_{{\rm bg}}\rangle}$ and ${\eta_{\text{res}}}$
is the resonant scattering phase shift given by the resonant K-matrix
${K_{\text{res}}=\tan\eta_{\text{res}}}$ of the Breit-Wigner form,
\begin{equation}
K_{\text{res}}=-\frac{\Gamma/2}{E-E_{m}-\Delta}.\label{eq:BreitWigner}
\end{equation}
In the limit of low energy ${E=\frac{\hbar^{2}k^{2}}{2\mu}}$, the
scattering length ${a=-\lim_{k\to0}\tan\eta/k}$ is therefore 
\begin{equation}
a=a_{{\rm bg}}-\frac{\lim_{k\to0}\Gamma/2k}{E_{m}+\Delta}.\label{eq:ScatteringLength}
\end{equation}
The scattering length ${a_{{\rm bg}}}$, the width $\Gamma$, and
shift $\Delta$ are thus the parameters that characterise the Fano-Feshbach
resonance at low energy. In the rest of this paper, we consider $\Gamma$
and $\Delta$ in the limit of low scattering energy.

The isolated resonance approximation is valid in the limit of small
coupling ${V_{{\rm co}}}$ with respect to level spacings in the closed
channel, so that effectively the resonant molecular level is well
isolated from the other levels. Indeed, the condition ${|\langle\Psi_{m}\vert\Psi_{{\rm c}}\rangle|\gg|\langle\Psi_{n}\vert\Psi_{{\rm c}}\rangle|}$
needed to ensure that ${|\Psi_{{\rm c}}\rangle}$ is approximately
proportional to ${|\Psi_{m}\rangle}$ gives the requirement 
\begin{equation}
\vert\tilde{E}_{m}-\tilde{E}_{n}\vert\gg\pi\vert\langle\Psi_{n}\vert V_{{\rm co}}\vert\Psi_{{\rm bg}}\rangle\langle\Psi_{m}\vert V_{{\rm co}}\vert\Psi_{{\rm bg}}\rangle\vert,\label{eq:small-coupling}
\end{equation}
where ${\tilde{E}_{m}=E_{m}+\langle\Psi_{m}|V_{{\rm co}}G_{0}^{+}V_{{\rm oc}}|\Psi_{m}\rangle}$
and ${\tilde{E}_{n}=E_{n}+\langle\Psi_{n}|V_{{\rm co}}G_{0}^{+}V_{{\rm oc}}|\Psi_{m}\rangle}$
are the dressed energies of the closed-channel molecular levels. We
call the regime where the inequality in Eq.~\eqref{eq:small-coupling}
is satisfied the diabatic limit. 

Even for large couplings $V_{co}$, it may be possible to apply the
isolated resonance in another basis for which the new coupling becomes
small. One such basis is the adiabatic basis that diagonalises at
each separation ${R}$ the potential matrix ${V_{ij}}$. The resulting
equations are formally similar to the original equations, where the
potentials ${V_{{\rm oo}}}$ and ${V_{{\rm cc}}}$ are replaced by
the adiabatic potentials ${V_{{\rm oo}}^{\prime}}$ and ${V_{{\rm cc}}^{\prime}}$,
and the couplings ${V_{{\rm oc}}}$ and ${V_{{\rm co}}}$ are replaced
by radial couplings ${V_{{\rm oc}}^{\prime}=-V_{{\rm co}}^{\prime}}$
of the form (see Appendix~1)
\begin{equation}
V_{{\rm oc}}^{\prime}(R)=-\frac{\hbar^{2}}{2\mu}\left[2\frac{Q(R)}{R}\frac{d}{dR}(R\cdot)+\frac{dQ(R)}{dR}\right],\label{eq:RadialCoupling}
\end{equation}
where the function ${Q(R)}$ in Eq.~\eqref{eq:RadialCoupling} is
given by 
\begin{equation}
Q(R)=-\frac{1}{2}\frac{d}{dR}\left[\arctan\left(\frac{2V_{{\rm oc}}(R)}{V_{{\rm oo}}(R)-V_{{\rm cc}}(R)}\right)\right].\label{eq:Qadiabatic}
\end{equation}
If the coupling $V_{{\rm oc}}^{\prime}$ happens to be small enough,
the condition (\ref{eq:small-coupling}) written in the new basis
may be satisfied and the isolated resonance approximation may be applied
again with a bound state ${|\Psi_{m}^{\prime}\rangle}$ among the
family of bound states ${|\Psi_{n}'\rangle}$ in the new closed channel.
We call this regime of weak adiabatic coupling the adiabatic limit.

\section{General dependence on ${a_{{\rm bg}}}$}

\label{sec:4}

We first consider the dependence of the width upon the background
scattering length for a vanishing colliding energy. Due to the isotropic
character of the inter-channel coupling, only the s-wave component
of the background scattering state contributes in Eq.~\eqref{eq:Width}.
At zero scattering energy, the s-wave component ${[\Psi_{{\rm bg}}(R)]_{s}}$
of the background state ${\vert\Psi_{{\rm bg}}\rangle}$ can be written
in terms of radial functions as 
\begin{equation}
\left[\Psi_{{\rm bg}}(\bm{R})\right]_{{\rm s}}\equiv\int\frac{d\Omega_{\bm{R}}}{4\pi}\langle\bm{R}|\Psi_{{\rm bg}}\rangle\propto\frac{u_{0}(R)-a_{{\rm bg}}u_{\infty}(R)}{R},\label{eq:Psibg-s-wave-component}
\end{equation}
where the integration over the solid angle ${\Omega_{\bm{R}}}$ selects
the s-wave component, and the radial functions ${u_{0}}$ and ${u_{\infty}}$
are two independent solutions of the open-channel radial equation,
\begin{equation}
\left(-\frac{\hbar^{2}}{2\mu}\frac{d^{2}}{dR^{2}}+V_{{\rm oo}}(R)\right)u(R)=0\label{eq:OpenChannel-radial-equation}
\end{equation}
with the asymptotic boundary conditions ${u_{0}(R)\xrightarrow[R\to\infty]{}R}$
and ${u_{\infty}(R)\xrightarrow[R\to\infty]{}1}$. The linear combination
of these two functions in Eq.~(\ref{eq:Psibg-s-wave-component})
corresponds precisely to the physical solution of (\ref{eq:OpenChannel-radial-equation})
that is regular at the origin. It is then clear from Eq.~(\ref{eq:Width})
and (\ref{eq:Psibg-s-wave-component}) that the width ${\Gamma}$
is the square of a quantity varying linearly with ${a_{{\rm bg}}}$.
In particular, for some value of ${a_{{\rm bg}}}$, the width ${\Gamma}$
vanishes.

Second, we examine the dependence of the shift of the resonance as
a function of the background scattering length. For this purpose,
we use the Green's function of the s-wave radial Schrödinger equation
for the open channel,
\begin{equation}
\left(-\frac{\hbar^{2}}{2\mu}\frac{d^{2}}{dR^{2}}+V_{{\rm oo}}(R)-E\right){\mathcal{G}}_{{\rm o}}^{E}(R,R^{\prime})=-\delta(R-R^{\prime}).
\end{equation}
It is related to the resolvent by 
\begin{equation}
{\mathcal{G}}_{{\rm o}}^{E}(R,R^{\prime})=4\pi RR'\left[G_{{\rm o}}(\bm{R},\bm{R}^{\prime})\right]_{s}\equiv RR'\int d\Omega_{\bm{R}}\langle\bm{R}|G_{{\rm o}}^{+}|\bm{R}'\rangle.\label{eq:GreensFunction-s-wave-component}
\end{equation}
In the following, we will focus on the low-energy regime. In this
regime, the Green's function ${\mathcal{G}_{{\rm o}}^{E}(R,R^{\prime})}$
is well approximated at short distances ${R,R^{\prime}\ll k^{-1}}$
by its zero-energy limit, 
\begin{multline}
{\mathcal{G}}_{{\rm o}}^{0}(R,R^{\prime})=\\
-\frac{2\mu}{\hbar^{2}}\begin{cases}
\left(u_{0}(R)-a_{{\rm bg}}u_{\infty}(R)\right)u_{\infty}(R^{\prime}) & \ \text{for }R<R^{\prime}\\
\left(u_{0}(R^{\prime})-a_{{\rm bg}}u_{\infty}(R^{\prime})\right)u_{\infty}(R) & \ \text{for }R>R^{\prime}
\end{cases}.\label{eq:RadialGreensFunction}
\end{multline}
Using this last expression, it follows from Eq.~(\ref{eq:Shift})
that the shift $\Delta$ varies linearly with ${a_{{\rm bg}}}$.

\section{Case of van der Waals interactions}

\label{sec:5}

Neutral atoms in their ground state interact via interactions that
decay as ${-C_{6}/R^{6}}$ (van der Waals potential) beyond a certain
radius $R_{0}$. In this case, one can give the explicit dependence
of the width and shift on ${a_{{\rm bg}}}$. The van der Waals tail
introduces a natural length scale ${R_{{\rm vdW}}}$ (or energy ${E_{{\rm vdW}}}$)
denoted as the van der Waals length (or energy): 
\begin{equation}
R_{{\rm vdW}}=\frac{1}{2}\left(\frac{2\mu C_{6}}{\hbar^{2}}\right)^{1/4}\ ;\ E_{{\rm vdW}}=\frac{\hbar^{2}}{2\mu R_{{\rm vdW}}^{2}}.\label{eq:vanderWaalsLength}
\end{equation}
In what follows, we will also use the Gribakin-Flambaum mean scattering
length ${\bar{a}=4\pi/\Gamma(1/4)^{2}R_{{\rm vdW}}}$ where ${\Gamma(\cdot)}$
denotes the Gamma function, giving ${\bar{a}\approx0.955978...R_{{\rm vdW}}}$
\cite{Gribakin1993}. The radial functions ${u_{0}(R)}$ and ${u_{\infty}(R)}$
are known analytically in the region ${R>R_{0}}$ of the van der Waals
tail: 
\begin{align}
u_{0}(R)/R_{{\rm vdW}} & =\sqrt{x}\Gamma(3/4)J_{-1/4}(2x^{-2})\label{eq:u0}\\
u_{\infty}(R) & =\sqrt{x}\Gamma(5/4)J_{1/4}(2x^{-2}),\label{eq:uinf}
\end{align}
where ${x=R/R_{{\rm vdW}}}$ and $J$ denotes the Bessel function.
In practice, ${R_{0}<R_{{\rm vdW}}}$ and in the short-range region
${R_{0}<R\lesssim R_{{\rm vdW}}}$, the functions exhibit rapid oscillations
that are well approximated by the semi-classical formulas, 
\begin{align}
u_{0}(R)/R_{{\rm vdW}} & \approx\Gamma(3/4)\frac{x^{3/2}}{\sqrt{\pi}}\cos(2x^{-2}-\pi/8)\label{eq:u0Approx}\\
u_{\infty}(R) & \approx\Gamma(5/4)\frac{x^{3/2}}{\sqrt{\pi}}\cos(2x^{-2}-3\pi/8).\label{eq:uinfApprox}
\end{align}
One deduces from Eq.~\eqref{eq:Width} and the normalization factor
of the scattering state ${\bar{\Psi}_{{\rm o}}^{E}}$, that the resonance
width vanishes at zero energy with a linear law in the colliding momentum
${k=\sqrt{2\mu E}/\hbar}$. 
Thus, in the limit of small ${k}$, one finds the explicit dependence
of $\Gamma$ and $\Delta$ upon ${a_{{\rm bg}}}$: 
\begin{align}
\frac{\Gamma}{2k\bar{a}} & =\pi E_{{\rm vdW}}\times\left|\sqrt{2}A-r_{{\rm bg}}B\right|^{2}\label{eq:WidthGeneral}\\
\Delta & =-\pi E_{{\rm vdW}}\times\Bigg[\sqrt{2}C-r_{{\rm bg}}B^{2}\Bigg],\label{eq:ShiftGeneral}
\end{align}
where we introduced the reduced background scattering length ${r_{{\rm bg}}=a_{{\rm bg}}/\bar{a}}$,
and the coefficients 
\begin{align}
A & =\int_{0}^{\infty}dxw(x)x^{3/2}\cos(2x^{-2}-\pi/8)\label{eq:A-B}\\
B & =\int_{0}^{\infty}dxw(x)x^{3/2}\cos(2x^{-2}-3\pi/8)\nonumber 
\end{align}
\begin{align}
C & =\int_{0}^{\infty}dxw(x)x^{3/2}\cos(2x^{-2}-3\pi/8)\label{eq:C}\\
 & \quad\times\int_{0}^{x}dx^{\prime}\cos(2x^{\prime-2}-\pi/8)w(x^{\prime})x^{\prime3/2}\nonumber \\
 & \quad+\int_{0}^{\infty}dxw(x)x^{3/2}\cos(2x^{-2}-\pi/8)\nonumber \\
 & \quad\times\int_{x}^{\infty}dx^{\prime}\cos(2x^{\prime-2}-3\pi/8)w(x^{\prime})x^{\prime3/2}\nonumber 
\end{align}
with ${w(x)=\sqrt{R_{{\rm vdW}}}E_{{\rm vdW}}^{-1}R\,W(R)}$ and ${W(R)=\left[V_{{\rm oc}}\Psi_{m}(\bm{R})\right]_{s}}$.

\subsection{Optical Fano-Feshbach resonance}

In the case of an optical Fano-Feshbach resonance, the closed-channel
potential ${V_{{\rm cc}}(R)}$ typically decays as ${V_{\text{cc}}(\infty)-C_{3}/R^{3}}$
(for pairs of alkali atoms in the ${S-P}$ electronic state). As a
result, the molecular state ${\Psi_{m}}$ is usually localised near
the Condon point ${R_{{\rm c}}}$ \cite{Bohn1999}, so that one can
make the approximation ${w(x)\approx w_{{\rm c}}\delta(x-x_{{\rm c}})}$,
with the obvious notation ${x_{{\rm c}}=R_{{\rm c}}/R_{{\rm vdW}}}$.
This gives: 
\begin{align}
A & =w_{{\rm c}}x_{{\rm c}}^{3/2}\cos(2x_{{\rm c}}^{-2}-\pi/8)\label{eq:Aoptical}\\
B & =w_{{\rm c}}x_{{\rm c}}^{3/2}\cos(2x_{{\rm c}}^{-2}-3\pi/8)\label{eq:Boptical}\\
C & =AB.\label{eq:Coptical}
\end{align}
Therefore 
\begin{align}
 & \frac{\Gamma}{2k\bar{a}}=\pi E_{{\rm vdW}}\times\left|\sqrt{2}A-r_{{\rm bg}}B\right|^{2}\label{eq:WidthOptical}\\
 & \Delta=-\pi E_{{\rm vdW}}\times\Bigg[\sqrt{2}A-r_{{\rm bg}}B\Bigg]B.\label{eq:ShiftOptical}
\end{align}
These formulae are akin to equations (3.6) and (3.7) in Ref.~\cite{Bohn1999}.
This gives a simple relation between $\Delta$ and ${\Gamma/2k\bar{a}}$:
\begin{equation}
\Delta=\frac{\Gamma}{2k\bar{a}}\times\left(r_{{\rm bg}}+\tan(2x_{{\rm c}}^{-2}-3\pi/8)-1\right)^{-1}.\label{eq:RelationOptical}
\end{equation}
This relation holds as long as ${R_{{\rm c}}\ll R_{{\rm vdW}}}$.
For larger Condon points, one has to use the general forms (\ref{eq:u0})
and (\ref{eq:uinf}) of ${u_{0}}$ and ${u_{\infty}}$, which gives
\begin{multline}
\boxed{\Delta=\frac{\Gamma}{2k\bar{a}}\times\left(r_{{\rm bg}}-\sqrt{2}\frac{J_{-1/4}(2x_{{\rm c}}^{-2})}{J_{1/4}(2x_{{\rm c}}^{-2})}\right)^{-1}}\\
\xrightarrow[r_{{\rm c}}\gg R_{{\rm vdW}}]{}\frac{\Gamma}{2k}\times(a_{{\rm bg}}-R_{{\rm c}})^{-1}.\label{eq:RelationOptical2}
\end{multline}

\subsection{Magnetic Fano-Feshbach resonance}

In the case of magnetic Fano-Feshbach resonances, the closed-channel
potential ${V_{{\rm cc}}(R)}$ has the same van der Waals tail as
the open-channel potential, i.e. $V_{\text{cc}}(R)=V_{\text{cc}}(\infty)-C_{6}/R^{6}$
for ${R>R_{0}}$. We assume that the molecular state involved in the
resonance is not too deeply bound in the closed channel, such that
its probability density is significant in the van der Waals region
$R>R_{0}$. This means that its binding energy $E_{b}=\vert E_{m}-V_{\text{cc}}(\infty)\vert$
is much smaller than $C_{6}/R_{0}^{6}$. In practice, $R_{0}\sim0.4\,R_{\text{vdW}}$,
which limits our consideration to $E_{b}\ll4000\,E_{\text{vdW}}$,
i.e. typically the last or next-to-last molecular level of the closed-channel
potential~\cite{Chin2010}. This situation is often the case in practice.
Indeed, the molecular state binding energy $E_{b}$ must be close
to the energy separation $\vert V_{\text{oo}}(\infty)-V_{\text{cc}}(\infty)\vert$
between the two channel thresholds. This separation results from Zeeman
and hyperfine splittings which are at most a few GHz for typical magnetic
fields less than $1000$~G. Since $E_{\text{vdW}}$ typically ranges
from 2 to 600~MHz for alkali atoms~\cite{Chin2010}, the condition
$E_{b}\lesssim4000\,E_{\text{vdW}}$ is often satisfied. 

In the interval of radii ${[R_{0},\sim\text{min}(1/\kappa,\;R_{\text{vdW}})]}$
where $\kappa=\sqrt{2\mu E_{b}/\hbar}$, the closed-channel potential
is well approximated by the van der Waals tail and the shape of the
molecular wave function is nearly energy-independent. In this interval,
the molecular wave function ${\Psi_{m}(R)=\langle\bm{R}|\Psi_{m}\rangle}$
may be approximated by the following zero-energy formula, similar
to Eq.~(\ref{eq:Psibg-s-wave-component}), 
\begin{equation}
\Psi_{m}(R)\propto\frac{u_{0}(R)-a_{{\rm c}}u_{\infty}(R)}{R},\label{eq:psim-zero-energy}
\end{equation}
where the radial functions $u_{0}(R)$ and $u_{\infty}(R)$ are given
in this interval by the semi-classical formulas in Eqs.~(\ref{eq:u0Approx})
and (\ref{eq:uinfApprox}). In Eq.~(\ref{eq:psim-zero-energy}) we
have introduced the length ${a_{{\rm c}}}$ that sets the phase in
the semi-classical region where the wave function of the bound state
oscillates. In the interval considered and in the small energy limit,
all the eigenfunctions of the closed channel have the same shape and
thus, in analogy with Eq.~(\ref{eq:Psibg-s-wave-component}) for
the open channel, we call ${a_{{\rm c}}}$ the \textit{closed-channel
scattering length}. It is, in general, different from the open-channel
scattering length ${a_{{\rm bg}}}$. 

In what follows, we make the additional assumption that the inter-channel
coupling can be neglected beyond a certain radius ${R_{{\rm free}}}$
satisfying the condition 
\begin{equation}
R_{0}<R_{{\rm free}}<\min(1/\kappa,\,R_{{\rm vdW}}),
\end{equation}
which is usually the case for magnetic resonances. As we shall see,
the crucial point is that the wave functions admit several oscillations
between $R_{0}$ and ${R_{{\rm free}}}$. Let us now consider the
adiabatic and diabatic limits.

\subsubsection{Adiabatic limit}

In the adiabatic basis, the inter-channel coupling is given by the
radial coupling ${V_{{\rm co}}^{\prime}}$ of Eq.~(\ref{eq:RadialCoupling}).
Therefore we have
\begin{align}
W(R) & =-\frac{\hbar^{2}}{2\mu R}\left[\frac{dQ}{dR}+2Q\frac{d}{dR}\right](R\Psi_{m}(R))\label{eq:w}\\
 & \approx-\frac{\hbar^{2}}{\mu R}Q(R)\frac{d}{dR}(R\Psi_{m}(R)).\nonumber 
\end{align}
We assume that the function ${W(R)}$ takes negligible values for
radii less than ${R_{0}}$ and that it is localised in a region where
the formula Eq.~(\ref{eq:psim-zero-energy}) is often a good approximation
for the molecular state ${\Psi_{m}}$~\cite{Mies1996}. It follows
that for ${{R_{0}}/{R_{{\rm vdW}}}<x<{R_{{\rm free}}}/{R_{{\rm vdW}}}}$
\footnote{Here, we have neglected the terms $\propto x^{1/2}$ with respect
to those $\propto x^{-3/2}$.}, 
\begin{multline}
w(x)=-\mathcal{W}(x)x^{-3/2}\times\\
\Big(\sqrt{2}\sin(2x^{-2}-\pi/8)-r_{{\rm c}}\sin(2x^{-2}-3\pi/8)\Big)\label{eq:wAdiabatic}
\end{multline}
where ${\mathcal{W}(x)=\lambda_{m}R_{{\rm vdW}}Q(R)}$ and ${\lambda_{m}}$
is a dimensionless normalisation factor depending on the molecular
wave function ${\Psi_{m}}$. We assume that ${\mathcal{W}(x)}$ has
a support that comprises several oscillations of ${\Psi_{m}(R)}$
and is varying slowly with respect to these oscillations. Replacing
the expression of Eq.~(\ref{eq:wAdiabatic}) into Eqs.~(\ref{eq:A-B},\ref{eq:C}),
and neglecting the terms with fast oscillations, one finds: 
\begin{align}
\frac{\Gamma}{2k\bar{a}} & =E_{{\rm vdW}}\times\frac{\pi}{4}\overline{\mathcal{W}}^{2}\left|r_{{\rm c}}-r_{{\rm bg}}\right|^{2}\label{eq:WidthAdiabatic}\\
\Delta & =-E_{{\rm vdW}}\times\frac{\pi}{4}\overline{\mathcal{W}}^{2}\left(r_{{\rm c}}-r_{{\rm bg}}\right),\label{eq:ShiftAdiabatic}
\end{align}
where ${r_{{\rm c}}=a_{{\rm c}}/\bar{a}}$ and ${r_{{\rm bg}}=a_{{\rm bg}}/\bar{a}}$
and 
\begin{equation}
\overline{\mathcal{W}}=\int_{R_{0}/R_{{\rm vdW}}}^{R_{{\rm free}}/R_{{\rm vdW}}}dx\,\mathcal{W}(x)=\lambda_{m}\int_{R_{0}}^{R_{{\rm free}}}dR\,Q(R).
\end{equation}
These expressions are consistent with the fact that the width and
shift vanish when the scattering lengths of the open and closed channels
are the same. Indeed, in the coupling region, both the open- and closed-channel
wave functions have the same short-range oscillations with the same
phase, and since the radial coupling operator shifts the phase of
one of them by ${\pi/2}$ through the derivative ${d/dR}$, the resulting
overlap is zero. From Eqs.~(\ref{eq:WidthAdiabatic}-\ref{eq:ShiftAdiabatic}),
we obtain the low-energy relation between the width and the shift:
\begin{equation}
\boxed{\Delta=\frac{\Gamma}{2k\bar{a}}\times(r_{{\rm bg}}-r_{{\rm c}})^{-1}}.\label{eq:RatioAdiabatic}
\end{equation}
This simple relation constitutes the main result of this paper. We
note in passing that it has a form similar to the relation obtained
for optical resonances - see Eq.~(\ref{eq:RelationOptical2}).

\subsubsection{Diabatic limit}

In the diabatic basis, the inter-channel coupling ${V_{{\rm co}}}$
is typically proportional to the exchange energy, i.e. the difference
between the triplet and singlet potentials for alkali atoms, which
decays exponentially with atomic separation. It is therefore localised
at separations smaller than the van der Waals length, in a region
that usually depends on the short-range details of the potentials.
There is therefore no obvious simplification from the formulas (\ref{eq:WidthGeneral})
and (\ref{eq:ShiftGeneral}) in general.

\subsection{Comparison with other works}

Our previous results, in particular Eq.~(\ref{eq:RatioAdiabatic}),
are inconsistent with formula (37) of Ref.~\cite{Chin2010}, which
reads as\footnote{We note that there is a global minus sign missing in Eq.~(37) of
Ref.~\cite{Chin2010}} 
\begin{equation}
\Delta=\frac{\Gamma}{2k\bar{a}}\times\frac{r_{{\rm bg}}-1}{1+(r_{{\rm bg}}-1)^{2}}.\label{eq:PaulFormula}
\end{equation}

Other works~\cite{Tiecke2010,Kokkelmans2014} have provided expressions
of $\Delta$ and $\Gamma/2k$ (see Eqs.~1.47 and 1.48 of Ref.~\cite{Kokkelmans2014})
that lead to
\begin{equation}
\Delta=\frac{\Gamma}{2k}\times(a_{bg}-r_{0})^{-1}\label{eq:ServaasFormula}
\end{equation}
where $r_{0}$ is a length scale associated with the range of the
open-channel interaction, i.e. typically of the order of $\bar{a}$.

Most strikingly, none of the above formulas depend on the closed channel,
unlike Eq.~(\ref{eq:RatioAdiabatic}) which depends on $a_{\text{c}}$.
The formula of Eq.~(\ref{eq:ServaasFormula}) was derived under the
approximation that the low-energy scattering properties of the open
channel are dominated by a pole (bound state) near its threshold,
neglecting contributions from other poles in the Mittag-Leffler expansion
of the resolvent $G_{\text{o}}$. This approximation seems to be valid
only for large $a_{\text{bg}}$, and one can check that in this limit,
both Eq.~(\ref{eq:ServaasFormula}) and our result Eq.~(\ref{eq:RatioAdiabatic})
indeed tend to the same limit. On the other hand, the formula of Eq.~(\ref{eq:PaulFormula})
is supposed to be valid for any value of $a_{\text{bg}}$ and without
any particular assumption on the closed channel. It was first published
in Eq.~(32) of Ref.~\cite{Goral2004}, and stated to be derived
from the MQDT. To understand the discrepancy with our result, we now
treat the two-channel resonance problem using the MQDT. 

\section{Multi-channel quantum defect theory}

\label{sec:6} We present here a self-contained derivation of the
MQDT, following the approach of Refs.~\cite{Mies1984a,Ruzic2013}.

\subsection{MQDT setup}

\subsubsection{Reference functions and short-range Y-matrix}

The coupled radial equations for the s-wave component of Eqs.~(\ref{eq:CC1}-\ref{eq:CC2})
read as follows, 
\begin{align}
 & \left(-\frac{\hbar^{2}}{2\mu}\frac{d^{2}}{dR^{2}}+V_{{\rm oo}}(R)-E\right)\psi_{{\rm o}}(R)+V_{{\rm oc}}(R)\psi_{{\rm c}}(R)=0\label{eq:CoupledRadial1}\\
 & \left(-\frac{\hbar^{2}}{2\mu}\frac{d^{2}}{dR^{2}}+V_{{\rm cc}}(R)-E\right)\psi_{{\rm c}}(R)+V_{{\rm co}}(R)\psi_{{\rm o}}(R)=0,\label{eq:CoupledRadial2}
\end{align}
where ${\psi_{{\rm o}}(R)=R[\Psi_{{\rm o}}(\bm{R})]_{s}}$ and ${\psi_{{\rm c}}(R)=R[\Psi_{{\rm c}}(\bm{R})]_{s}}$
are the s-wave radial wave functions. The starting point of MQDT is
that the channels are uncoupled for radii ${R>R_{{\rm free}}}$. In
this region, one can express the two independent solutions ${\psi^{(1)}=(\psi_{{\rm o}}^{(1)},\psi_{{\rm c}}^{(1)})}$
and ${\psi^{(2)}=(\psi_{{\rm o}}^{(2)},\psi_{{\rm c}}^{(2)})}$ of
Eqs.~\eqref{eq:CoupledRadial1},\eqref{eq:CoupledRadial2}, as linear
combinations of reference functions (${\hat{f}_{{\rm o}},\hat{g}_{{\rm o}}}$)
and ${(\hat{f}_{{\rm c}},\hat{g}_{{\rm c}})}$, which are solutions
of the diagonal potentials ${V_{{\rm oo}}}$ and ${V_{{\rm cc}}}$
in each channel at energy $E$: 
\begin{equation}
\left(\begin{array}{cc}
\psi_{{\rm o}}^{(1)} & \psi_{{\rm o}}^{(2)}\\
\psi_{{\rm c}}^{(1)} & \psi_{{\rm c}}^{(2)}
\end{array}\right)=\left(\begin{array}{cc}
\hat{f}_{{\rm o}}-\hat{g}_{{\rm o}}Y_{{\rm oo}} & \qquad-\hat{g}_{{\rm o}}Y_{{\rm oc}}\\
-\hat{g}_{{\rm c}}Y_{{\rm co}} & \qquad\hat{f}_{{\rm c}}-\hat{g}_{{\rm c}}Y_{{\rm cc}}
\end{array}\right)\label{eq:MatrixY}
\end{equation}
The functions ${\hat{f}_{{\rm o}}}$ and ${\hat{f}_{{\rm c}}}$ are
taken to be regular at the origin, i.e. they vanish at ${R=0}$, and
therefore the functions ${\hat{g}_{{\rm o}}}$ and ${\hat{g}_{{\rm c}}}$
must be irregular. They are normalised such that the Wronskians $W[\hat{f}_{{\rm o}},\hat{g}_{{\rm o}}]=\hat{f}_{{\rm o}}\hat{g}_{{\rm o}}'-\hat{f}_{{\rm o}}'\hat{g}_{{\rm o}}=1$
and ${W[\hat{f}_{{\rm c}},\hat{g}_{{\rm c}}]=1}$. One finds in the
limit of weak coupling (see Appendix 2), 
\begin{align}
Y_{{\rm co}} & =-(\hat{f}_{{\rm c}}\vert\hat{f}_{{\rm o}})\label{eq:Yco}\\
Y_{{\rm oc}} & =-(\hat{f}_{{\rm o}}\vert\hat{f}_{{\rm c}})=Y_{{\rm co}}^{*}\label{eq:Yoc}\\
Y_{{\rm oo}} & =-\Big(\hat{f}_{{\rm o}}\Big\vert\hat{g}_{{\rm c}}(\hat{f}_{{\rm c}}\vert\hat{f}_{{\rm o}})_{<}\Big)-\Big(\hat{f}_{{\rm o}}\Big\vert\hat{f}_{{\rm c}}(\hat{g}_{{\rm c}}\vert\hat{f}_{{\rm o}})_{>}\Big)\label{eq:Yoo}\\
Y_{{\rm cc}} & =-\Big(\hat{f}_{{\rm c}}\Big\vert\hat{g}_{{\rm o}}(\hat{f}_{{\rm o}}\vert\hat{f}_{{\rm c}})_{<}\Big)-\Big(\hat{f}_{{\rm c}}\Big\vert\hat{f}_{{\rm o}}(\hat{g}_{{\rm o}}\vert\hat{f}_{{\rm c}})_{>}\Big),\label{eq:Ycc}
\end{align}
where we have introduced the short-hand notations 
\begin{align}
(\hat{f}_{i}\vert\hat{g}_{j}) & \equiv\int_{0}^{\infty}dR\hat{f}_{i}(R)\frac{2\mu}{\hbar^{2}}V_{ij}(R)\hat{g}_{j}(R)\label{eq:fg}\\
(\hat{f}_{i}\vert\hat{g}_{j})_{<} & \equiv\int_{0}^{R}dR^{\prime}\hat{f}_{i}(R^{\prime})\frac{2\mu}{\hbar^{2}}V_{ij}(R^{\prime})\hat{g}_{j}(R^{\prime})\label{eq:fginf}\\
(\hat{f}_{i}\vert\hat{g}_{j})_{>} & \equiv\int_{R}^{\infty}dR^{\prime}\hat{f}_{i}(R^{\prime})\frac{2\mu}{\hbar^{2}}V_{ij}(R^{\prime})\hat{g}_{j}(R^{\prime}).\label{eq:fgsup}
\end{align}
The second ingredient of MQDT is that in the uncoupled region the
reference functions are usually governed by the tails of the potentials
${V_{{\rm oo}}}$ and ${V_{{\rm cc}}}$. For example, assuming that
the potential ${V_{{\rm oo}}}$ has a van der Waals tail with van
der Waals length ${R_{{\rm vdW}}}$, the reference functions ${\hat{f}_{{\rm o}}}$
and ${\hat{g}_{{\rm o}}}$ may be written in the region ${R_{0}<R\lesssim R_{{\rm vdW}}}$:
\begin{align}
\hat{f} & _{{\rm o}}\approx R_{{\rm vdW}}^{1/2}\,\frac{1}{2}x^{3/2}\sin\left(\frac{2}{x^{2}}+\frac{\pi}{8}+\varphi_{{\rm o}}\right)\label{eq:fhato}\\
\hat{g}_{{\rm o}} & \approx R_{{\rm vdW}}^{1/2}\,\frac{1}{2}x^{3/2}\cos\left(\frac{2}{x^{2}}+\frac{\pi}{8}+\varphi_{{\rm o}}\right),\label{eq:ghato}
\end{align}
which are two independent linear combinations of Eqs.~(\ref{eq:u0Approx}-\ref{eq:uinfApprox}).
The phase ${\varphi_{{\rm o}}}$ is adjusted to make the function
$\hat{f}_{{\rm o}}$ regular at the origin. The above expressions
do not depend on the energy $E$, because the potentials are deep
enough in the interval ${[R_{0},R_{{\rm vdW}}]}$ that wave functions
are nearly energy-independent there. On the other hand, the asymptotic
part (i.e. for ${R\gg R_{{\rm vdW}}}$) of the functions ${(\hat{f}_{{\rm o}},\hat{g}_{{\rm o}})}$
{[}respectively ${(\hat{f}_{{\rm c}},\hat{g}_{{\rm c}})}${]} is a
linear combination of free-wave solutions in the open channel (respectively
closed channel) and is strongly energy dependent. 

\subsubsection{Elimination of the closed channel}

The reference functions ${\hat{f}_{{\rm c}}}$ and ${\hat{g}_{{\rm c}}}$
being in the closed channel, they are in general exponentially divergent
at large distance. Only one particular linear combination of ${\psi^{(1)}}$
and ${\psi^{(2)}}$ is the physical solution of Eqs.~\eqref{eq:CoupledRadial1}
and \eqref{eq:CoupledRadial2}, having a non-diverging component in
the closed channel ${\psi_{{\rm c}}\propto\exp(-\kappa R)}$ for large
$R$, where ${\kappa=\sqrt{V_{{\rm cc}}(\infty)-E}}$. We define ${\cot\gamma_{{\rm c}}}$
such that 
\begin{equation}
\hat{f}_{{\rm c}}+\cot\gamma_{{\rm c}}\hat{g}_{{\rm c}}\propto\exp(-\kappa R)\quad\text{for}\quad R\to\infty.\label{eq:Tangammac}
\end{equation}
Therefore, we must have ${\psi_{{\rm c}}\propto\hat{f}_{{\rm c}}+\cot\gamma_{{\rm c}}\hat{g}_{{\rm c}}}$,
which implies that for ${R>R_{{\rm free}}}$,
\begin{equation}
\psi_{{\rm o}}\propto\hat{f}_{{\rm o}}-\underbrace{\left(Y_{{\rm oo}}-Y_{{\rm oc}}(Y_{{\rm cc}}+\cot\gamma_{{\rm c}})^{-1}Y_{{\rm co}}\right)}_{\tilde{Y}}\hat{g}_{{\rm o}}.\label{eq:Ytilde}
\end{equation}

\subsubsection{Energy-normalised reference functions}

In the open channel, one can define another set of reference functions
${f_{{\rm o}}}$ and ${g_{{\rm o}}}$ that are energy-normalised solutions
of the potential ${V_{{\rm oo}}}$, such that 
\begin{align}
f_{{\rm o}} & \xrightarrow[R\to\infty]{}\sqrt{\frac{2\mu}{4\pi^{2}\hbar^{2}k}}\sin(kR+\eta_{{\rm o}})\label{eq:fo}\\
g_{{\rm o}} & \xrightarrow[R\to\infty]{}-\sqrt{\frac{2\mu}{4\pi^{2}\hbar^{2}k}}\cos(kR+\eta_{{\rm o}}).\label{eq:go}
\end{align}
Again ${f_{{\rm o}}}$ is chosen to be regular ${f_{{\rm o}}(0)=0}$,
so that the phase shift ${\eta_{{\rm o}}\xrightarrow[k\to0]{}-ka_{{\rm o}}}$
is simply the physical phase shift of the potential ${V_{{\rm oo}}}$.
The function ${f_{{\rm o}}}$ is thus the radial function of the s-wave
component of the energy-normalised scattering state ${\vert\bar{\Psi}_{{\rm o}}^{E}\rangle}$.

One can connect the reference functions ${\hat{f}_{{\rm o}}}$ and
${\hat{g}_{{\rm o}}}$ to the functions ${f_{{\rm o}}}$ and ${g_{{\rm o}}}$
as follows: 
\begin{align}
f_{{\rm o}} & =\sqrt{\frac{2\mu}{4\pi^{2}\hbar^{2}}}C^{-1}\hat{f}_{{\rm o}}\label{eq:foexp}\\
g_{{\rm o}} & =\sqrt{\frac{2\mu}{4\pi^{2}\hbar^{2}}}C(\mathcal{G}\hat{f}_{{\rm o}}+\hat{g}_{{\rm o}}),\label{eq:goexp}
\end{align}
provided that the short-range phase ${\varphi_{{\rm o}}}$ is adjusted
to satisfy 
\begin{equation}
\tan\varphi_{{\rm o}}=\frac{1}{1-r_{{\rm o}}}\quad\text{with }r_{{\rm o}}\equiv\frac{a_{{\rm o}}}{\bar{a}}.\label{eq:tanphio}
\end{equation}
Then, using the zero-energy analytical solutions (\ref{eq:u0}-\ref{eq:uinf})
of the van der Waals problem (which are also valid at low energy for
${R\ll k^{-1}}$), one finds for small $k$, 
\begin{align}
 & C^{-1}\;\underset{k\to0}{\approx}\;\sqrt{k\bar{a}(1+(1-r_{{\rm o}})^{2})}\label{eq:C-1}\\
 & \mathcal{G}\;\underset{k\to0}{\approx}\;r_{{\rm o}}-1.\label{eq:G}
\end{align}

\subsubsection{$K$-matrix resulting from the inter channel coupling}

Expressing the open-channel radial wave function ${\psi_{{\rm o}}}$
of Eq.~\eqref{eq:Ytilde} in terms of the reference functions ${f_{{\rm o}}}$
and ${g_{{\rm o}}}$ in Eqs.~(\ref{eq:foexp},\ref{eq:goexp}) gives,
for ${R>R_{{\rm free}}}$, 
\begin{equation}
\psi_{{\rm o}}\propto f_{{\rm o}}-\tilde{Y}(1+\tilde{Y}\mathcal{G})^{-1}C^{-2}g_{{\rm o}}.\label{eq:psio}
\end{equation}
Then, one can directly identify the $K$-matrix, ${\tilde{K}=\tan\tilde{\eta}}$
resulting from the coupling of the open channel with the closed channel:
\begin{equation}
\tilde{K}=\tilde{Y}(1+\tilde{Y}\mathcal{G})^{-1}C^{-2}.
\end{equation}
Indeed, one can check from Eqs.~\eqref{eq:fo} and \eqref{eq:psio}
that the total phase shift is 
\begin{equation}
\eta=\eta_{{\rm o}}+\tilde{\eta}\label{eq:TotalPhaseShift}
\end{equation}
and a resonance occurs for ${\eta=\pi/2+n\pi}$, i.e. at a pole of
${K=\tan\eta}$. Using Eq.~(\ref{eq:Ytilde}), the explicit form
of ${\tilde{K}}$ reads 
\begin{equation}
\tilde{K}=\frac{C^{-2}}{\left(Y_{{\rm oo}}-Y_{{\rm oc}}(Y_{{\rm cc}}+\cot\gamma_{{\rm c}})^{-1}Y_{{\rm co}}\right)^{-1}+\mathcal{G}}.\label{eq:Kres}
\end{equation}

\subsection{Weak-coupling limit}

\subsubsection{MQDT formulas}

For weak coupling, the pole of $K$ appears for a scattering energy
$E$ near the energy ${E_{m}}$ of a molecular level in the potential
${V_{{\rm cc}}}$. Let us consider a scattering energy $E$ that is
close to the energy ${E_{m}}$. By definition of a bound state, when
$E$ is exactly equal to ${E_{m}}$ the coefficient ${\cot\gamma_{{\rm c}}}$
must be equal to zero such that the combination ${\hat{f}_{{\rm c}}(R)+\cot\gamma_{{\rm c}}\hat{g}_{{\rm c}}(R)}$,
which converges at ${R\to\infty}$, is also regular at ${R=0}$. We
denote this bound-state radial wave function by $\hat{f}_{m}(R)$.
When $E$ is close to but different from ${E_{m}}$, one can make
the Taylor expansion, 
\begin{equation}
\cot\gamma_{{\rm c}}\sim\alpha(E-E_{m})\quad{\rm with}\quad\alpha=\left[\frac{d(\cot\gamma_{{\rm c}})}{dE}\right]_{E=E_{m}}.\label{eq:TaylorExpansion}
\end{equation}
The coefficient ${\alpha}$ in Eq.~\eqref{eq:TaylorExpansion} is
related to the normalisation of the bound-state wave function $\hat{f}_{m}$
- see Appendix 3. In this approximation, one gets from Eq.~(\ref{eq:Kres}),
\begin{equation}
\tilde{K}\approx\frac{C^{-2}}{\left(Y_{{\rm oo}}-Y_{{\rm oc}}(Y_{{\rm cc}}+\alpha(E-E_{m}))^{-1}Y_{{\rm co}}\right)^{-1}+\mathcal{G}}.\label{eq:KtildeNearResonance}
\end{equation}
When ${E}$ is sufficiently far from ${E_{m}}$, then ${\tilde{K}\approx\tilde{K}_{{\rm o}}}$
with ${\tilde{K}_{{\rm o}}=\tan\tilde{\eta}_{{\rm o}}=C^{-2}Y_{{\rm oo}}/(1+\mathcal{G}Y_{{\rm oo}})}$.
One can then rewrite ${\tilde{K}}$ in the form 
\begin{equation}
\tilde{K}=\frac{\tilde{K}_{{\rm o}}+K_{\text{res}}}{1-\tilde{K}_{{\rm o}}K_{\text{res}}}\label{eq:KtildeWithK0andKres}
\end{equation}
i.e. 
\begin{equation}
\tilde{\eta}=\tilde{\eta}_{{\rm o}}+\eta_{\text{res}},\label{eq:etaTilde}
\end{equation}
where ${K_{\text{res}}=\tan\eta_{\text{res}}}$ has the standard Breit-Wigner
form for an isolated resonance given by Eq.~(\ref{eq:BreitWigner}),
with the width and shift, 
\begin{align}
\Gamma/2 & =C^{-2}\frac{1}{(1+\mathcal{G}Y_{{\rm oo}})^{2}+C^{-4}Y_{{\rm oo}}^{2}}\vert Y_{{\rm oc}}\vert^{2}\alpha^{-1}\label{eq:Gamma}\\
\Delta & =\left(\frac{\mathcal{G}(1+\mathcal{G}Y_{{\rm oo}})+C^{-4}Y_{{\rm oo}}}{(1+\mathcal{G}Y_{{\rm oo}})^{2}+C^{-4}Y_{{\rm oo}}^{2}}\vert Y_{{\rm oc}}\vert^{2}-Y_{{\rm cc}}\right)\alpha^{-1}.\label{eq:Delta}
\end{align}
Combining Eqs.~(\ref{eq:TotalPhaseShift}) and (\ref{eq:etaTilde}),
one retrieves the total phase shift of Eq.~(\ref{eq:ScatteringPhaseShift}),
\begin{equation}
\eta=\underbrace{\eta_{{\rm o}}+\tilde{\eta}_{{\rm o}}}_{\eta_{{\rm bg}}}+\eta_{\text{res}}.\label{eq:TotalPhaseShift2}
\end{equation}
At low scattering energy ${E}$, one retrieves the scattering length
of Eq.~(\ref{eq:ScatteringLength}) and, using Eqs.~(\ref{eq:C-1}-\ref{eq:G}),
one obtains 
\begin{align}
a_{{\rm bg}} & =-\lim_{k\to0}\frac{\eta_{{\rm o}}+\tilde{\eta}_{{\rm o}}}{k}=a_{{\rm o}}-\frac{\bar{a}(1+(r_{{\rm o}}-1)^{2})Y_{{\rm oo}}}{1+(r_{{\rm o}}-1)Y_{{\rm oo}}}\label{eq:abg}\\
\frac{\Gamma}{2k\bar{a}} & =\frac{\bar{a}(1+(r_{{\rm o}}-1)^{2})}{(1+(r_{{\rm o}}-1)Y_{{\rm oo}})^{2}}\vert Y_{{\rm oc}}\vert^{2}\alpha^{-1}\label{eq:Gamma0}\\
\Delta & =\left(\frac{r_{{\rm o}}-1}{1+(r_{{\rm o}}-1)Y_{{\rm oo}}}\vert Y_{{\rm oc}}\vert^{2}-Y_{{\rm cc}}\right)\alpha^{-1}.\label{eq:Delta0}
\end{align}
From the equations (\ref{eq:Yco}-\ref{eq:Ycc}), one can see that
the off-diagonal matrix elements ${Y_{{\rm oc}}}$ and ${Y_{{\rm co}}}$
of the short-range Y-matrix are of first order in the coupling ${V_{{\rm co}}}$,
whereas the diagonal elements ${Y_{{\rm oo}}}$ and ${Y_{{\rm cc}}}$
are of second order. Therefore, in the limit of weak coupling, one
may neglect ${Y_{{\rm oo}}}$ in the above expressions, resulting
in 
\begin{align}
a_{{\rm bg}} & \approx a_{{\rm o}}\label{eq:abgApprox}\\
\frac{\Gamma}{2k\bar{a}} & \approx\bar{a}(1+(r_{{\rm bg}}-1)^{2})\vert Y_{{\rm oc}}\vert^{2}\alpha^{-1}\label{eq:Gamma0Approx}\\
\Delta & \approx\left((r_{{\rm bg}}-1)\vert Y_{{\rm oc}}\vert^{2}-Y_{{\rm cc}}\right)\alpha^{-1}.\label{eq:Delta0Approx}
\end{align}
It may seem natural to neglect ${Y_{{\rm cc}}}$ as well. Indeed,
the formula of Eq.~(\ref{eq:PaulFormula}) was obtained from the
above equations by neglecting both diagonal elements ${Y_{{\rm oo}}}$
and ${Y_{{\rm cc}}}$, as can be checked easily. However, a closer
inspection of Eq.~(\ref{eq:Delta0Approx}) shows that both ${\vert Y_{{\rm oc}}\vert^{2}}$
and ${Y_{{\rm cc}}}$ are of second order in the coupling. One may
therefore not neglect ${Y_{{\rm cc}}}$ in that equation. In the next
subsection, we show that one can retrieve from Eqs.~(\ref{eq:Gamma0Approx}-\ref{eq:Delta0Approx})
the results of the isolated resonance theory, Eqs.~(\ref{eq:Shift}-\ref{eq:Width}),
provided ${Y_{{\rm cc}}}$ is not neglected.

\subsubsection{Equivalence with the isolated resonance approximation}

As shown in section \ref{sec:3} the isolated resonance approximation
consists in considering only one resonant molecular level and neglecting
the contribution from other molecular levels in the closed channel.
Similarly, in the MQDT formalism, we have made a Taylor expansion~\eqref{eq:TaylorExpansion}
near a particular molecular level. The contribution from other molecular
levels is represented by the matrix element ${Y_{{\rm oo}}}$. In
this section, we show that neglecting this term in MQDT is indeed
equivalent to the isolated resonance approximation, leading back to
Eqs.~(\ref{eq:Shift},\ref{eq:Width}). 

Let us first calculate the width of the resonance. Neglecting ${Y_{{\rm oo}}}$
in Eq.~(\ref{eq:Gamma}) and using Eq.~(\ref{eq:Yoc}) and (\ref{eq:foexp}),
one gets 
\begin{align}
\Gamma/2 & \approx C^{-2}\left|(\hat{f}_{{\rm o}}\vert\hat{f}_{{\rm c}})\right|^{2}\alpha^{-1}\label{eq:Gamma2}\\
 & =\left|\int_{0}^{\infty}dR\;\sqrt{\frac{4\pi^{2}\hbar^{2}}{2\mu}}f_{{\rm o}}(R)\;\frac{2\mu}{\hbar^{2}}V_{{\rm oc}}(R)\;\hat{f}_{{\rm c}}(R)\alpha^{-1/2}\right|^{2}\nonumber \\
 & =\pi\left|\int_{0}^{\infty}4\pi dR\;f_{{\rm o}}(R)\;V_{{\rm oc}}(R)\;\hat{f}_{{\rm c}}(R)\sqrt{\frac{2\mu}{4\pi\hbar^{2}\alpha}}\right|^{2}\nonumber \\
 & =\pi\left|\langle\bar{\Psi}_{{\rm o}}^{E}\vert V_{{\rm oc}}\vert\Psi_{{\rm c}}\rangle\right|^{2},\nonumber 
\end{align}
where ${\Psi_{{\rm c}}(R)=\frac{\hat{f}_{{\rm c}}(R)}{R}\sqrt{\frac{2\mu}{4\pi\hbar^{2}\alpha}}}$.
Close to the resonance, ${\Psi_{{\rm c}}}$ is simply the closed-channel
bound state ${\Psi_{m}}$ satisfying ${\langle\Psi_{m}\vert\Psi_{m}\rangle=1}$
(see details in Appendix 3). Hence, we retrieve the formula~(\ref{eq:Width})
for the width in the isolated resonance approximation.

Let us now calculate the shift of the resonance. Neglecting ${Y_{{\rm oo}}}$
in Eq.~(\ref{eq:Delta}) and using Eqs.~(\ref{eq:Yoc}) and (\ref{eq:Ycc}),
we get 
\begin{equation}
\Delta=\alpha^{-1}\left(\mathcal{G}\vert(\hat{f}_{{\rm c}}\vert\hat{f}_{{\rm o}})\vert^{2}+\left(\hat{f}_{{\rm c}}\Big\vert\hat{g}_{{\rm o}}(\hat{f}_{{\rm o}}\vert\hat{f}_{{\rm c}})_{<}+\hat{f}_{{\rm o}}(\hat{g}_{{\rm o}}\vert\hat{f}_{{\rm c}})_{>}\right)\right).
\end{equation}
Then, writing ${\vert(\hat{f}_{{\rm c}}\vert\hat{f}_{{\rm o}})\vert^{2}}$
as ${\left(\hat{f}_{{\rm c}}\Big\vert\hat{f}_{{\rm o}}\left((\hat{f}_{{\rm o}}\vert\hat{f}_{{\rm c}})_{<}+(\hat{f}_{{\rm o}}\vert\hat{f}_{{\rm c}})_{>}\right)\right)}$,
one finds 
\begin{equation}
\Delta=\alpha^{-1}\left(\hat{f}_{{\rm c}}\Big\vert(\mathcal{G}\hat{f}_{{\rm o}}+\hat{g}_{{\rm o}})(\hat{f}_{{\rm o}}\vert\hat{f}_{{\rm c}})_{<}+\hat{f}_{{\rm o}}(\mathcal{G}\hat{f}_{{\rm o}}+\hat{g}_{{\rm o}}\vert\hat{f}_{{\rm c}})_{>}\right).
\end{equation}
Using Eqs.~(\ref{eq:foexp}-\ref{eq:goexp}), we obtain 
\begin{equation}
\Delta=\frac{4\pi^{2}\hbar^{2}}{2\mu}\alpha^{-1}\left(\hat{f}_{{\rm c}}\Big\vert g_{{\rm o}}(f_{{\rm o}}\vert\hat{f}_{{\rm c}})_{<}+f_{{\rm o}}(g_{{\rm o}}\vert\hat{f}_{{\rm c}})_{>}\right).
\end{equation}
Finally, using ${f_{m}(R)=R\Psi_{m}(R)=\sqrt{\frac{2\mu}{4\pi\hbar^{2}\alpha}}\hat{f}_{{\rm c}}(R)}$,
we arrive at 
\begin{equation}
\Delta=16\pi^{3}\left(\frac{\hbar^{2}}{2\mu}\right)^{2}\left(f_{m}\Big\vert g_{{\rm o}}(f_{{\rm o}}\vert f_{m})_{<}+f_{{\rm o}}(g_{{\rm o}}\vert f_{m})_{>}\right)\label{eq:Delta-equiv}
\end{equation}
which is exactly the same as the isolated-resonance approximation
formula \eqref{eq:Shift} for the shift. Indeed, starting from Eq.~\eqref{eq:Shift},
one finds 
\begin{multline}
\Delta=4\pi\int RdR\int R^{\prime}dR^{\prime}\Psi_{m}(R)V_{{\rm co}}(R)\\
\times{\mathcal{G}}_{{\rm o}}^{E}(R,R^{\prime})V_{{\rm oc}}(R^{\prime})\Psi_{m}(R^{\prime})
\end{multline}
where the s-wave Green's function ${{\mathcal{G}}_{{\rm o}}^{E}(R,R^{\prime})}$
of Eq.~\eqref{eq:GreensFunction-s-wave-component} can be approximated
at low energy and in the range of the inter-channel coupling by its
zero-energy limit ${{\mathcal{G}}_{{\rm o}}^{0}(R,R^{\prime})}$.
By using Eq.~\eqref{eq:RadialGreensFunction} and the relations ${f_{{\rm o}}=\sqrt{\frac{2\mu k}{4\pi^{2}\hbar^{2}}}u_{0}}$
and ${g_{{\rm o}}=-\sqrt{\frac{2\mu}{4\pi^{2}\hbar^{2}k}}u_{\infty}}$
deduced from Eqs.~(\ref{eq:fo}-\ref{eq:go}), one obtains 
\begin{multline}
\Delta=16\pi^{3}\int_{0}^{\infty}dRf_{m}(R)V_{{\rm co}}(R)\\
\times\Bigg(\int_{0}^{R}dR^{\prime}f_{{\rm o}}(R^{\prime})g_{{\rm o}}(R)V_{{\rm oc}}(R^{\prime})\\
+\int_{R}^{\infty}dR^{\prime}f_{{\rm o}}(R)g_{{\rm o}}(R^{\prime})V_{{\rm oc}}(R^{\prime})\Bigg)f_{m}(R^{\prime})
\end{multline}
which is exactly Eq.~\eqref{eq:Delta-equiv}. This shows that the
isolated resonance approximation is equivalent to the MQDT in which
${Y_{{\rm oo}}}$ is neglected. We note that neglecting ${Y_{{\rm cc}}}$
in addition to ${Y_{{\rm oo}}}$ would lead to the erroneous result
${\Delta=\frac{4\pi\hbar^{2}}{2\mu}\mathcal{G}\vert(f_{m}\vert\hat{f}_{{\rm o}})\vert^{2}}$.

We conclude that our results are consistent with the MQDT, whereas
the formula (\ref{eq:PaulFormula}) should be discarded as resulting
from the generally invalid neglect of ${Y_{{\rm cc}}}$. Although
the formula (\ref{eq:PaulFormula}) was reported to be verified numerically
for various magnetic resonances, we surmise that it was done mostly
for resonances with a large background scattering length ${a_{{\rm bg}}}$,
for which the shift is conspicuous and can be more easily determined.
In that limit, both Eq.~(\ref{eq:PaulFormula}) and our result (\ref{eq:RatioAdiabatic})
reduce to ${\Delta\approx\Gamma/(2ka_{{\rm bg}})}$. This would explain
why the shortcomings of Eq.~(\ref{eq:PaulFormula}) have been so
far unnoticed. 

\section{Comparison with two-channel calculations}

\begin{figure}
\includegraphics[width=8cm]{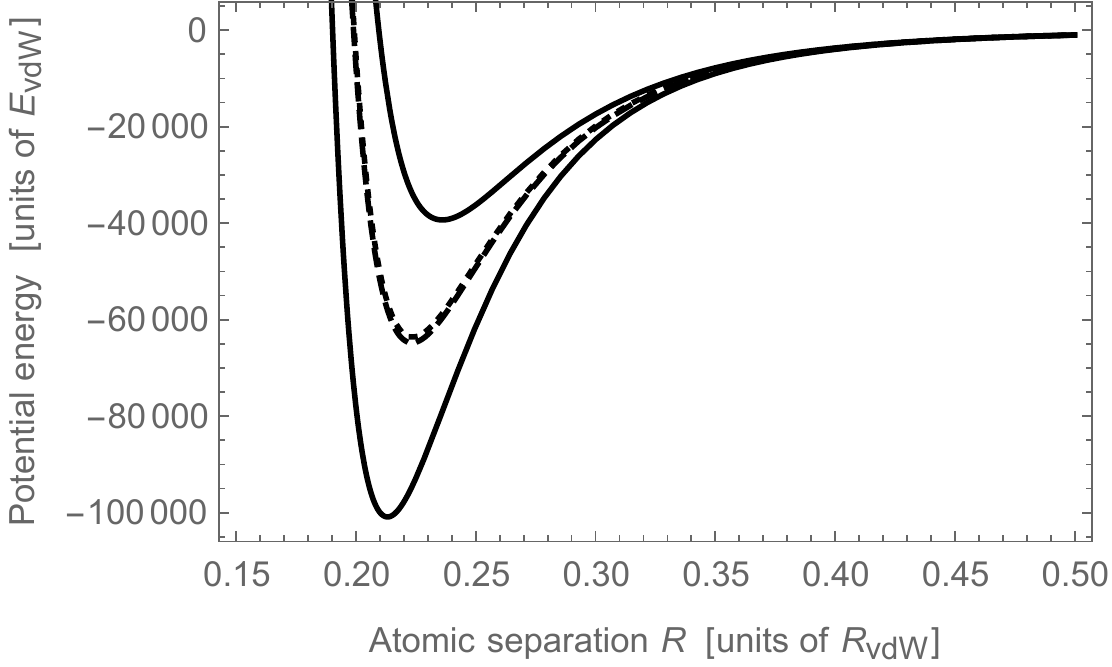}

\caption{\label{fig:Potentials}Adiabatic and diabatic potentials of the two-channel
model. The solid curves represent the potentials $V_{-}$ and $V_{+}$
of Eqs.~(\ref{eq:Vplus}-\ref{eq:Vminus}) for $\sigma_{-}=0.196\,R_{\text{vdW}}$,
$\sigma_{+}=0.200\,R_{\text{vdW}}$, and $\delta=0$. For a weak adiabatic
coupling strength $W=0.01$, the diabatic potentials $V_{\text{oo}}$
and $V_{\text{cc}}$ nearly coincide with the potentials $V_{+}$
and $V_{-}$. For a stronger adiabatic coupling strength $W=0.39$,
the diabatic potentials $V_{\text{oo}}$ and $V_{\text{cc}}$ , shown
by dashed and dotted curves, are nearly equal at short distance to
the average of the two adiabatic potentials $V_{-}$ and $V_{+}$.}

\end{figure}

We now compare our prediction given by Eq.~(\ref{eq:RatioAdiabatic})
to the result of numerical two-channel calculations. For this purpose,
we consider the following two-channel model defined in the adiabatic
basis. The two adiabatic potentials $V_{+}$ and $V_{-}$ (see Appendix
1) are constructed as Lennard-Jones potentials, i.e. the sum of a
van der Waals attraction and a $1/R^{12}$ repulsion modelling the
repulsive core, with an extra interaction decaying exponentially,
$\pm A\exp(-R/B)$, physically corresponding to an exchange interaction:

\begin{align}
V_{-}(R) & =16E_{\text{vdW}}\left[\frac{\sigma_{-}^{6}}{R^{12}}-\frac{R_{\text{vdW}}^{6}}{R^{6}}\right]-A\exp(-R/B)\label{eq:Vplus}\\
V_{+}(R) & =16E_{\text{vdW}}\left[\frac{\sigma_{+}^{6}}{R^{12}}-\frac{R_{\text{vdW}}^{6}}{R^{6}}\right]+A\exp(-R/B)+\delta\label{eq:Vminus}
\end{align}
where $\sigma_{-}$ and $\sigma_{+}$ are of the order of $0.2R_{\text{vdW}}^{2}$,
$A=2.0\times10^{7}E_{\text{vdW}}$, and $B=R_{\text{vdW}}/30$. The
energy separation $\delta$ between the two potentials at large distance
can be varied to account for the Zeeman effect of the magnetic field
on the potentials. The adiabatic coupling function $Q(R)$ is taken
as
\[
Q(R)=\frac{W}{L}\text{sech}^{2}\left(\frac{R-R_{0}}{L}\right)
\]
where $L=0.08\,R_{\text{vdW}}$ and $R_{0}=0.43R_{\text{vdW}}$ so
that it is located in the region of van der Waals oscillations of
the wave functions. This model thus constitutes a simplified yet realistic
description of magnetic Fano-Feshbach resonances. For sufficiently
small values of $W$, the adiabatic coupling is perturbative and the
model meets the conditions of derivation of the formula given by Eq.~(\ref{eq:RatioAdiabatic}).
For a small value such as $W=0.01$, the diabatic potential $V_{\text{oo}}$
and $V_{\text{cc}}$ {[}obtained by inverting Eq.~(\ref{eq:AdiabaticPotentialMatrix}){]}
are almost identical to the adiabatic potentials $V_{-}$ and $V_{+}$.
We consider values up to $W=0.39$, corresponding to a more realistic
situation where both diabatic potentials are nearly degenerate at
short distance with the average of the two adiabatic potentials $\frac{1}{2}(V_{-}+V_{+})$,
as shown in Fig.~\ref{fig:Potentials}.

\begin{figure}
\includegraphics[width=8cm]{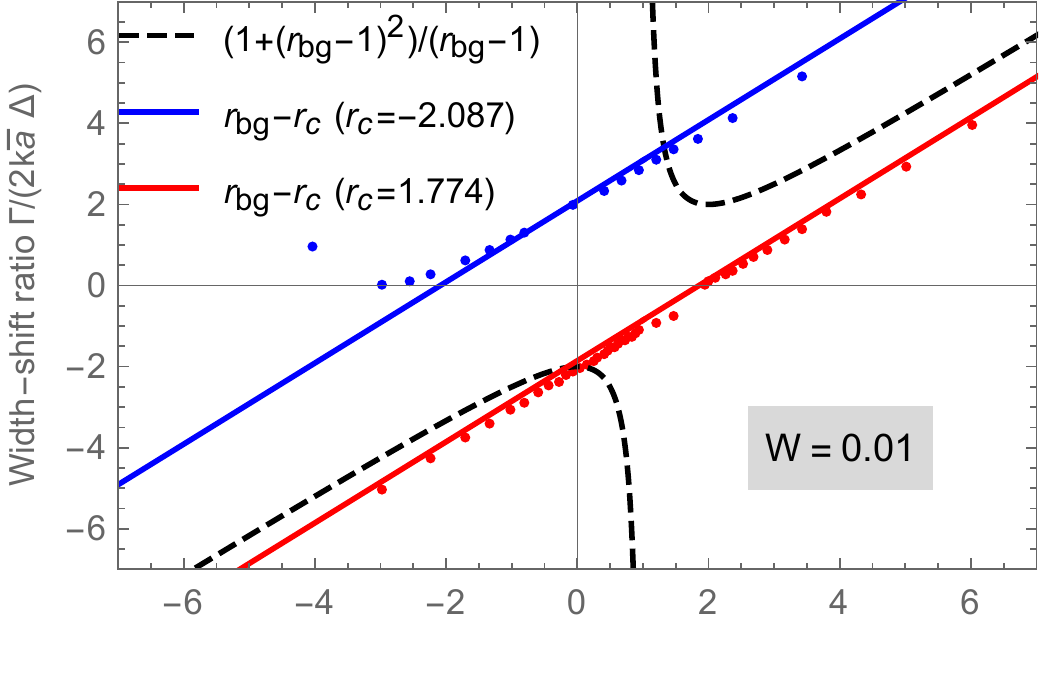}

\includegraphics[width=8cm]{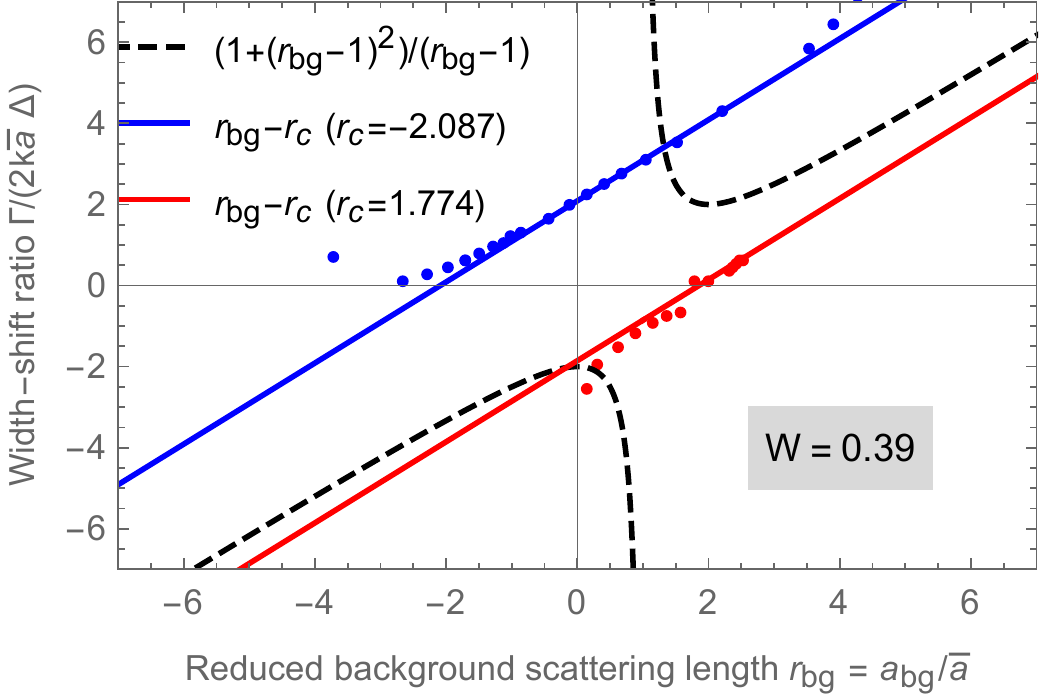}

\caption{\label{fig:TwoChannelModel}Ratio between the width parameter $\Gamma_{0}=\lim_{k\to0}\Gamma/2k\bar{a}$
and shift $\Delta$, for resonances with different values of the reduced
background scattering length $r_{\text{bg}}$ obtained by varying
the short-range parameter $\sigma_{-}$ of the open-channel adiabatic
potential $V_{-}$. The upper panel is obtained for a coupling strength
$W=0.01$ for which the diabatic potentials are almost equal to the
adiabatic potentials $V_{+}$ and $V_{-}$, and the lower panel is
obtained for a coupling strength $W=0.39$ for which the diabatic
potentials are nearly degenerate.}
\end{figure}

Changing $\sigma_{-}$ or $\sigma_{+}$ enables one to independently
control the scattering lengths $a_{\text{o}}$ and $a_{\text{c}}$,
as well as the positions of the bare energy levels of each adiabatic
potential $V_{-}$ or $V_{+}$ . These quantities are easily obtained
numerically using either a propagation method or a matrix representation
with finite differences. To solve the two coupled equations, we first
return to the original diabatic basis by using the transformation
matrix $P$ of Eq.~(\ref{eq:TransformationMatrix}) and then solve
the coupled equations (\ref{eq:CC1}-\ref{eq:CC2}) numerically. We
can obtain the scattering length $a$ as a function of the energy
separation $\delta$, which exhibits divergences that can be fitted
by the formula $a=a_{\text{bg}}-\Gamma_{0}\bar{a}/(\delta-E_{0}).$
Comparing with Eq.~(\ref{eq:ScatteringLength}), we obtain the width
of the corresponding resonance $\lim_{k\to0}\frac{\Gamma}{2k\bar{a}}=\Gamma_{0}$
and its shift $\Delta=E_{b}-E_{0}$, where $E_{b}$ is the binding
energy of the closest molecular level (of energy $E_{m}=\delta-E_{b}$)
in the closed-channel adiabatic potential $V_{+}$. We repeat this
procedure for different values of $a_{\text{o}}$ that produce different
values of $\Gamma_{0}$, $\Delta$ and $a_{\text{bg}}$. It is then
possible to plot the ratio $\Gamma_{0}/\Delta$ as a function of $a_{\text{bg}}$.
The result is shown in Fig.~\ref{fig:TwoChannelModel}, for two different
fixed values $a_{\text{c}}=-2.087\,R_{\text{vdW}}$ (blue points)
and $a_{\text{c}}=1.774\,R_{\text{vdW}}$ (red points), and for two
different values $W=0.01$ (upper panel) and $W=0.39$ (lower panel).
The analytical formula of Eq.~(\ref{eq:RatioAdiabatic}) predicts
a linear behaviour that crosses zero at $a_{\text{bg}}=a_{\text{c}}$
(red and blue lines), which is verified to a large extent by the numerical
data. In contrast, the formula of Eq.~(\ref{eq:PaulFormula}) does
not depend on $a_{\text{c}}$ and predicts a qualitatively different
behaviour (dashed curve) that is inconsistent with the numerical results.

We note that while the data for $W=0.01$ are well in the perturbative
regime, as we find that $a_{\text{bg}}\approx a_{\text{o}}$ to a
very good accuracy, the data for $W=0.39$ are at the limit of validity
of the perturbative regime, because $a_{\text{bg}}$ is only approximately
equal to $a_{\text{o}}$ and in a smaller range of values. Nevertheless,
within that range, the formula of Eq.~(\ref{eq:RatioAdiabatic})
appears to remain verified even at this coupling strength.

\section{Application to lithium-6}

\label{sec:7}
\begin{figure}
\hspace*{\fill}\includegraphics[viewport=30bp 0bp 430bp 516bp,width=8.5cm]{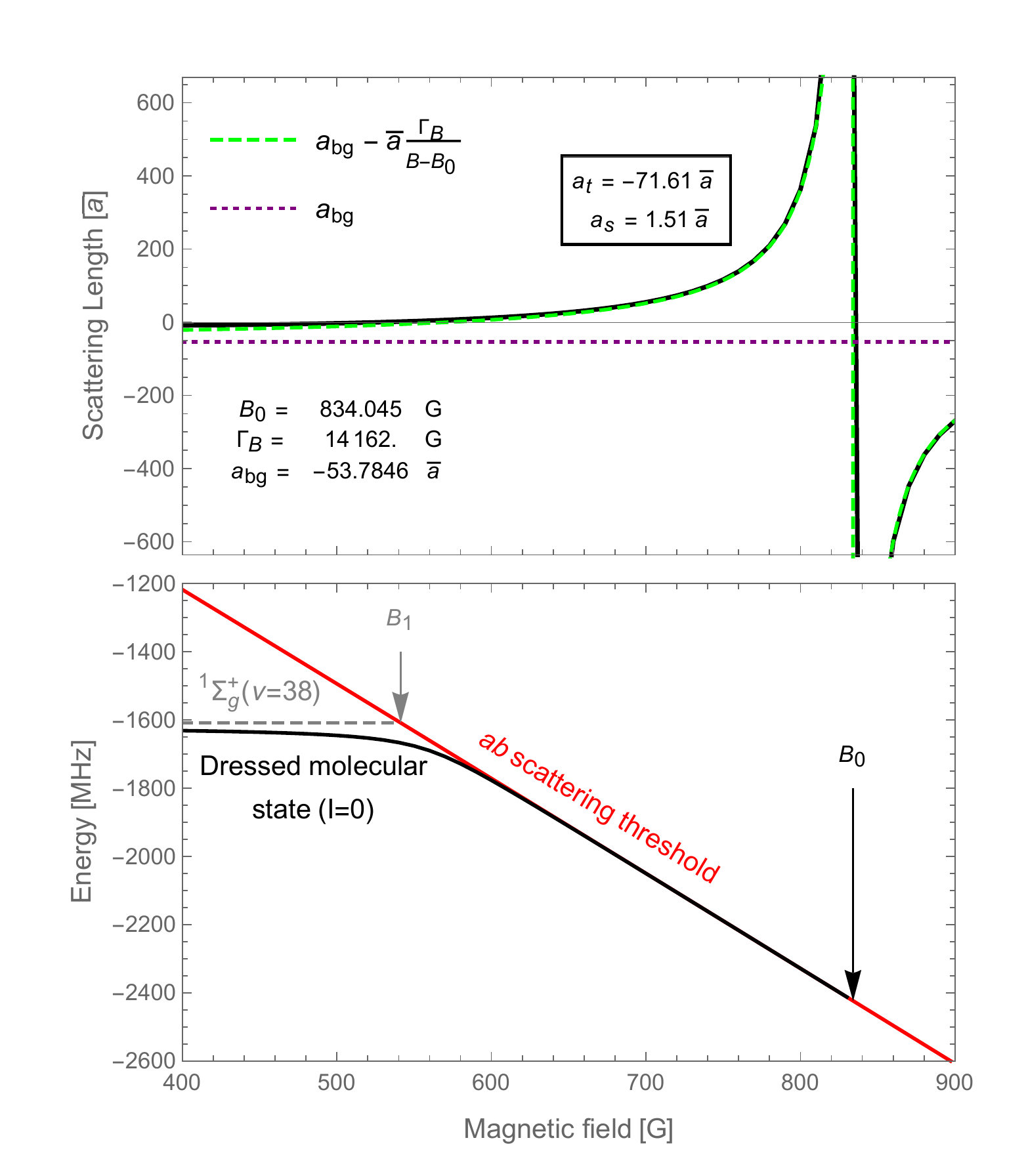}\hspace*{\fill}
\caption{\label{fig:Broad-Fano-Feshbach-resonance}Broad Fano-Feshbach resonance
of lithium-6 atoms in the hyperfine states ${a}$ and ${b}$ (first
and second lowest states) around a magnetic field intensity of 834~G.
Upper panel: s-wave scattering length as a function of the magnetic
field intensity. Lower panel: energy spectrum (below the $ab$ scattering
threshold) as a function of the magnetic field intensity. The solid
black curve represents the energy of the dressed molecular state (with
total nuclear spin ${I=0}$) associated with the broad resonance.
The black arrow shows the resonance position $B_{0}$ at which the
dressed molecular state reaches the threshold.\textcolor{red}{{} }The
dashed grey line shows the energy of the last level of the singlet
$^{1}\Sigma_{g}^{+}$ potential corresponding to the bare molecular
state causing the resonance, intersecting the threshold at the bare
resonance position $B_{1}$. Lengths are expressed in units of ${\bar{a}=1.5814}$~nm.}
\end{figure}
\begin{figure}
\centerline{\includegraphics[width=8cm]{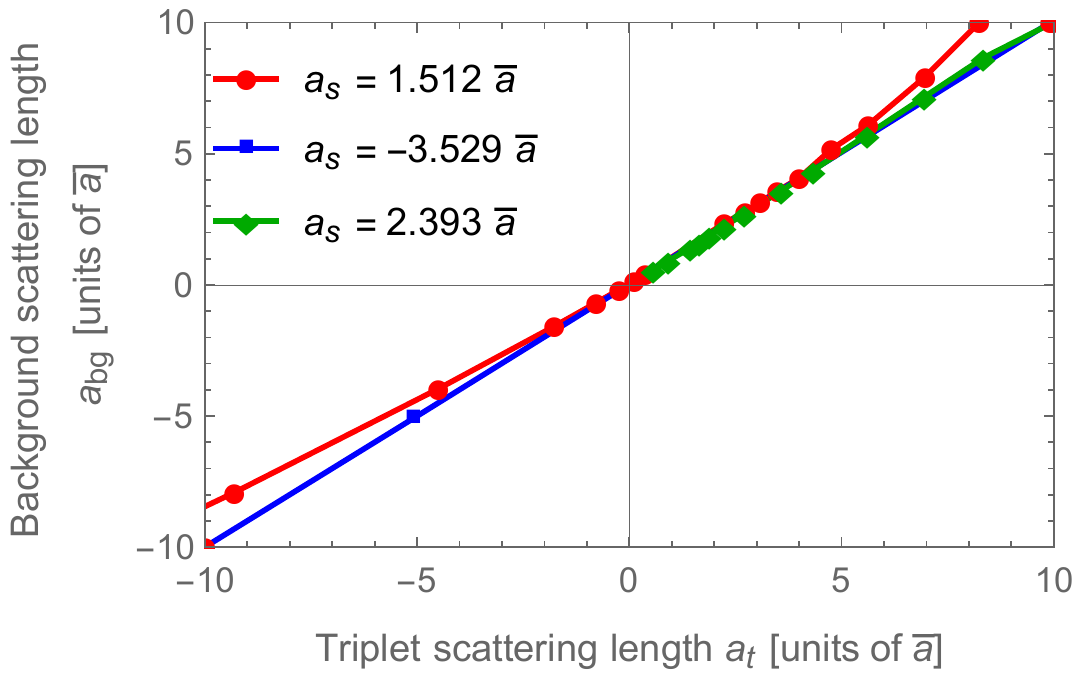}} \caption{\label{fig:abg-vs-at}Background scattering length ${a_{{\rm bg}}}$
of the lithium-6 resonance as a function of the triplet scattering
length ${a_{{\rm t}}}$, for different values of the singlet scattering
length ${a_{{\rm s}}}$. This graph shows that ${a_{{\rm bg}}}$ is
nearly independent of ${a_{{\rm s}}}$ and is approximately equal
to ${a_{{\rm t}}}$. Lengths are expressed in units of ${\bar{a}=1.5814}$~nm.}
\end{figure}

Although we were able in the previous section to check our formula,
Eq.~(\ref{eq:RatioAdiabatic}), by numerically solving the two-channel
equations (\ref{eq:CC1}-\ref{eq:CC2}) with van der Waals potentials,
it is more difficult to verify that formula from experimental data
or even from a realistic multi-channel calculation. While the width
and background scattering lengths can usually be determined both experimentally
and theoretically, the shift from the bare molecular state is more
ambiguous, because it is not directly observable if the coupling causing
the resonance cannot be tuned, as is the case for conventional magnetic
Fano-Feshbach resonances. 

Here, we consider the case of the broad resonance of lithium-6 atoms
in the two lowest hyperfine states near a magnetic field intensity
of 834 G, for which the bare molecular state causing the resonance
has been identified as the last vibrational level of the singlet potential~\cite{Simonucci2005,Chin2010}.
This is illustrated in Fig.~\ref{fig:Broad-Fano-Feshbach-resonance},
which was obtained from a realistic multi-channel calculation taking
into account the five relevant hyperfine channels. The upper panel
shows that the variation of the scattering length is well fitted by
the formula ${a=a_{{\rm bg}}-\bar{a}\Gamma_{B}/(B-B_{0})}$, making
it possible to determine the ``magnetic width'' of the resonance
${\Gamma_{B}=14162}$~G, the resonance position ${B_{0}=834.045}$~G,
and the background scattering length ${a_{{\rm bg}}=-53.78\,\bar{a}}$,
where ${\bar{a}=1.5814}$~nm. The lower panel of Fig.~\ref{fig:Broad-Fano-Feshbach-resonance}
shows the molecular energy, which reaches the threshold at the resonance
point ${B_{0}}$ (solid black curve, corresponding to a molecular
state with total nuclear spin ${I=0}$) and the energy $E_{m}$ of
the bare molecular state causing the resonance (dashed black line,
corresponding to the last level $\nu=38$ of the singlet potential),
which reaches the threshold at the magnetic field intensity $B_{1}=541.28$~G.

Comparing with Eq.~(\ref{eq:ScatteringLength}), one finds 
\begin{equation}
\delta\mu\,\Gamma_{B}=\lim_{k\to0}\frac{\Gamma}{2k\bar{a}}\quad;\quad\delta\mu(B_{1}-B_{0})=\lim_{k\to0}\Delta,\label{eq:magnetic_width-shift}
\end{equation}
where ${\delta\mu}$ is the difference of magnetic moments between
the bare molecule and the separated atoms. One can then calculate
the ratio of the two quantities in Eq.~\eqref{eq:magnetic_width-shift}:
\begin{equation}
\lim_{k\to0}\frac{\Gamma}{2k\bar{a}\Delta}=\frac{\Gamma_{B}}{B_{1}-B_{0}}\approx-48.38\label{eq:ratio_Delta-Gamma}
\end{equation}
This value turns out to compare well with the value ${(1+(r_{{\rm bg}}-1)^{2})(r_{{\rm bg}}-1)^{-1}=-54.79}$
given by formula (\ref{eq:PaulFormula}). However, as explained in
the previous section, this is because the value of ${r_{{\rm bg}}}$
is unusually large, so that the formula reduces to ${\approx r_{{\rm bg}}=-53.79}$,
which is also the limit of the formula given by Eq.~(\ref{eq:RatioAdiabatic}).
The broad 834~G resonance therefore does not allow one to discriminate
between these formulas. For this purpose, one needs to theoretically
change the value of the background scattering length.

It is not easy, in general, to control only the background scattering
length by altering the Hamiltonian of the system. However, the case
of lithium-6 is somewhat fortunate in that respect, because the background
scattering length turns out to be given essentially by the the triplet
scattering length ${a_{{\rm t}}}$ of the system, as shown in Fig.~\ref{fig:abg-vs-at},
while the closed channel is controlled by the singlet scattering length
${a_{{\rm s}}}$, since the close-channel bare molecular state is
of a singlet nature. These values can be changed by slightly altering
the shape of the triplet and singlet potentials at short distances.
\begin{figure*}
\includegraphics[height=10cm]{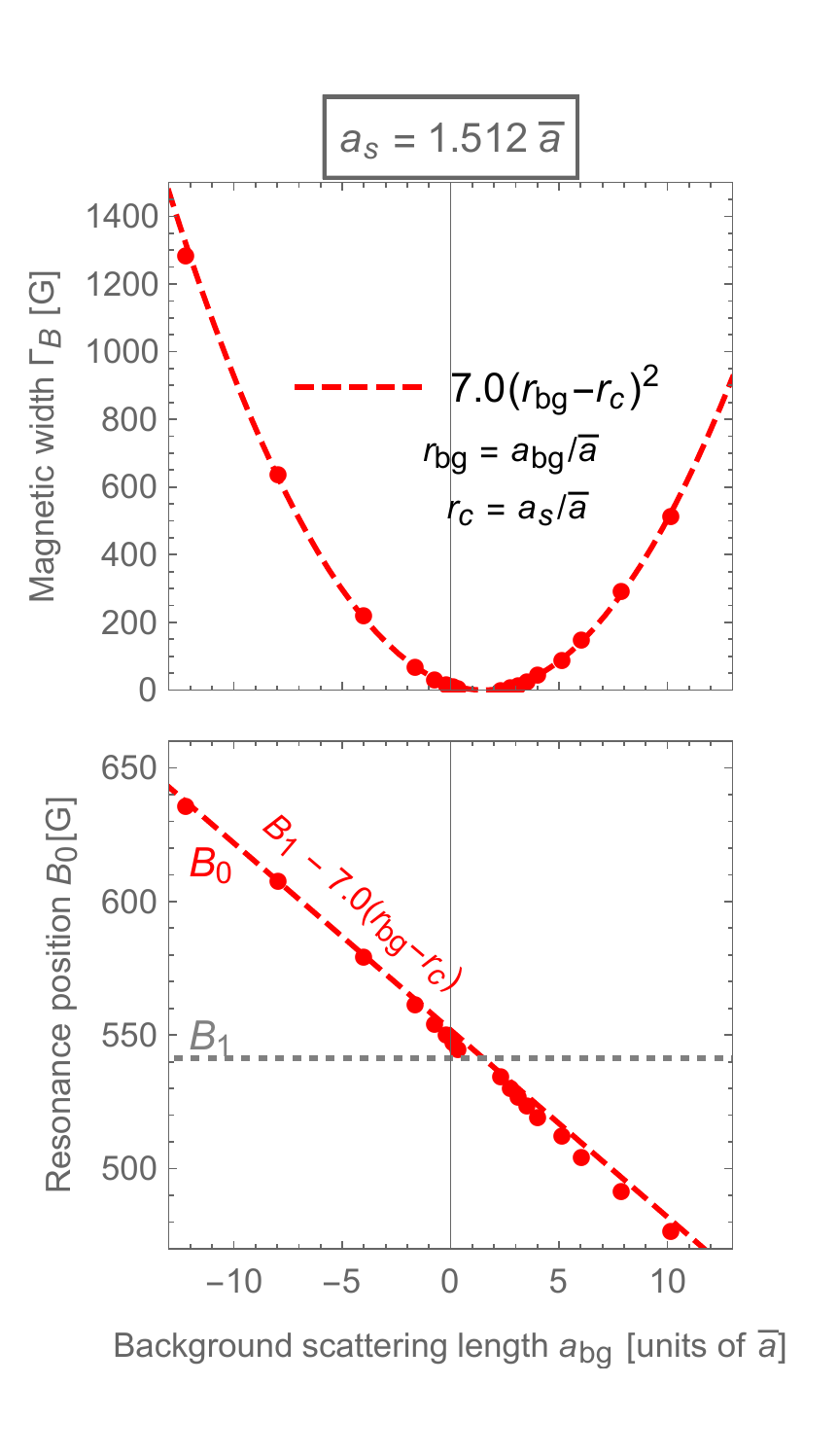}\includegraphics[height=10cm]{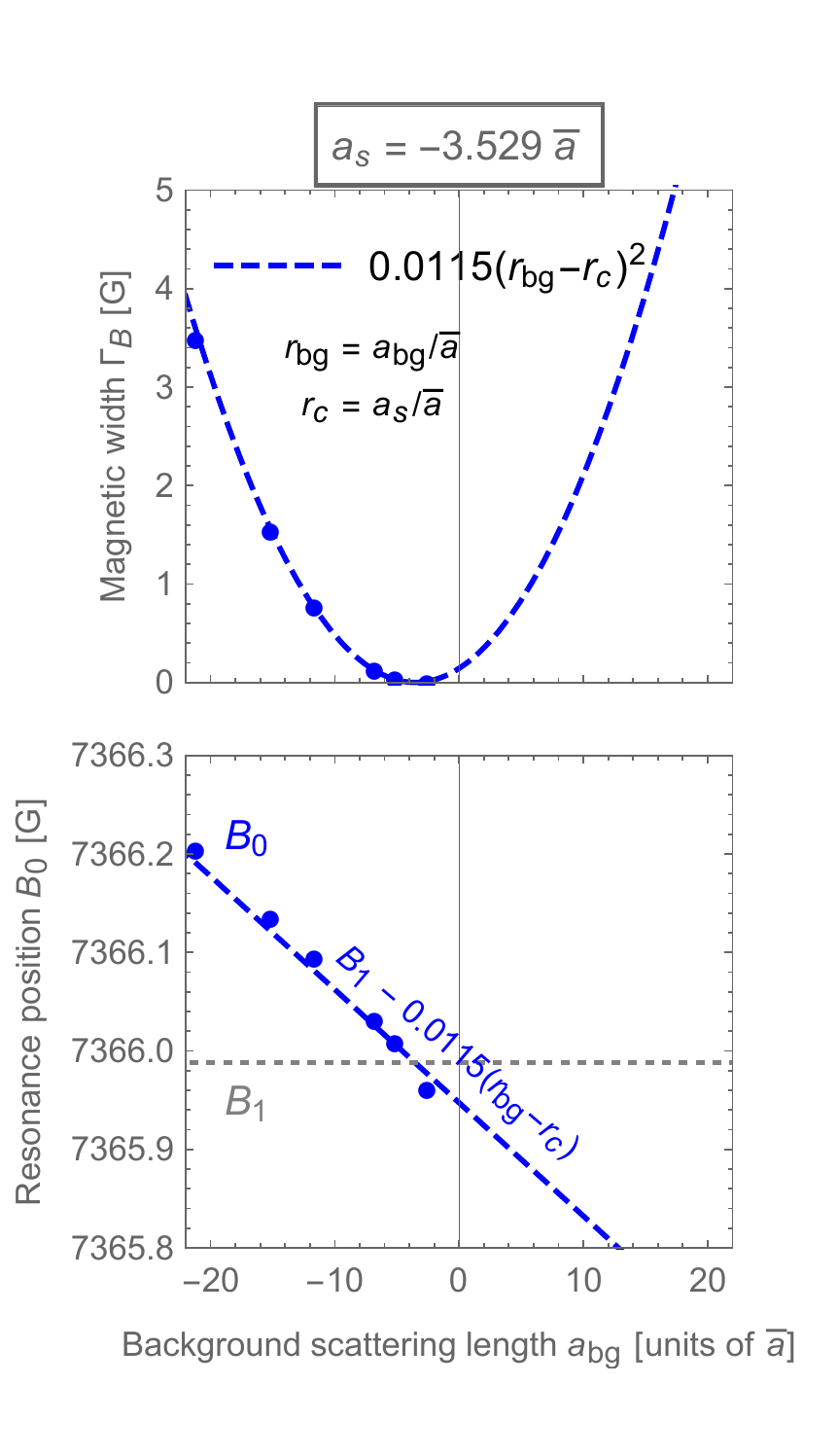}\includegraphics[height=10cm]{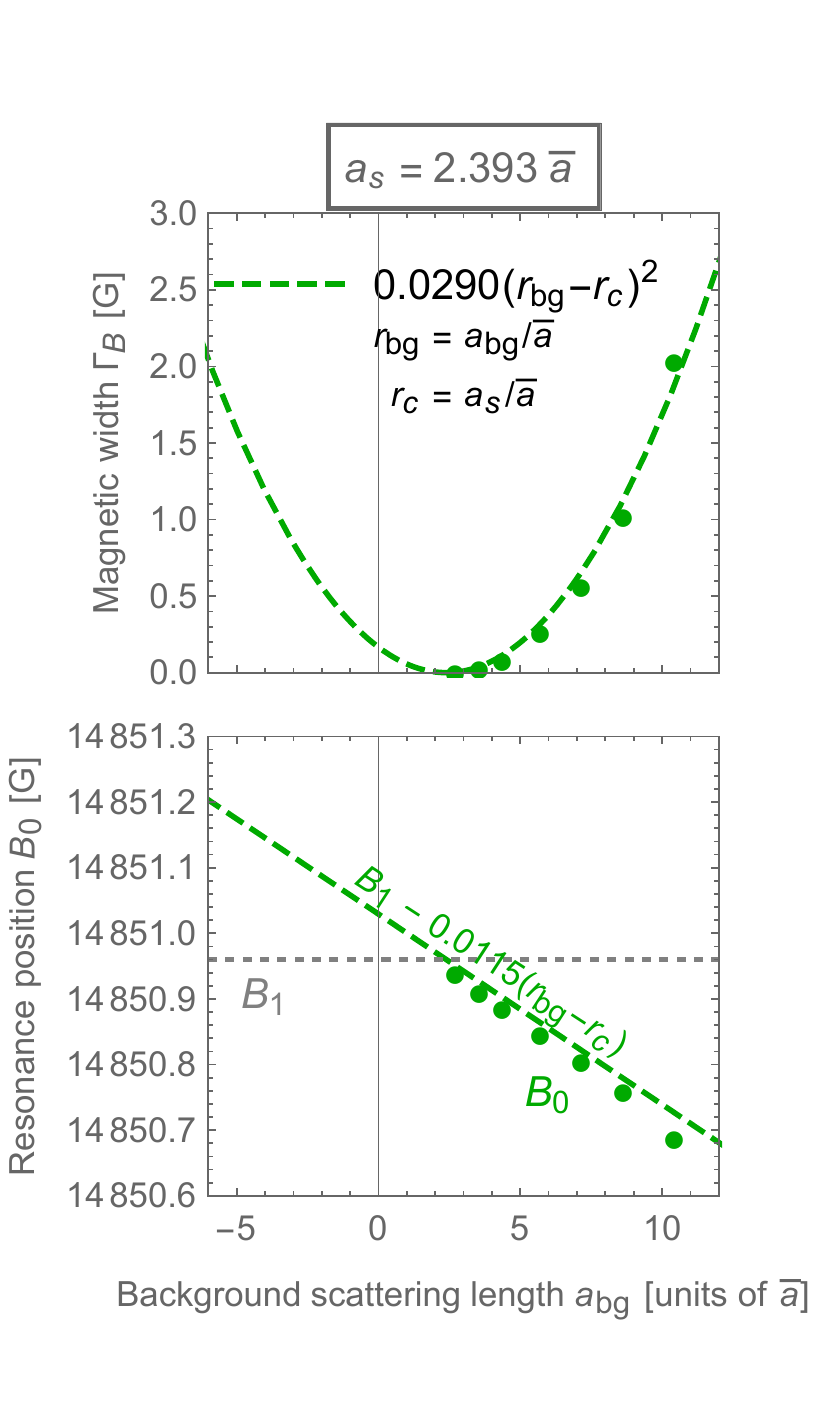}\caption{\label{fig:WidthAndPosition} Plots of the magnetic width ${\Gamma_{B}}$
and position $B_{0}$ of the resonance {[}see Eqs.~\eqref{eq:magnetic_width-shift},\eqref{eq:ratio_Delta-Gamma}
and discussion around{]} as a function of the reduced background scattering
length ${r_{{\rm bg}}=a_{{\rm bg}}/\bar{a}}$, for three different
singlet scattering lengths ${a_{{\rm s}}}$ indicated in the box at
the top of each figure. The dots are obtained from a multi-channel
calculation, while the dashed curves represent Eqs.~(\ref{eq:WidthAdiabatic})
and (\ref{eq:ShiftAdiabatic}).}
\end{figure*}

\begin{figure}
\includegraphics[width=8.5cm]{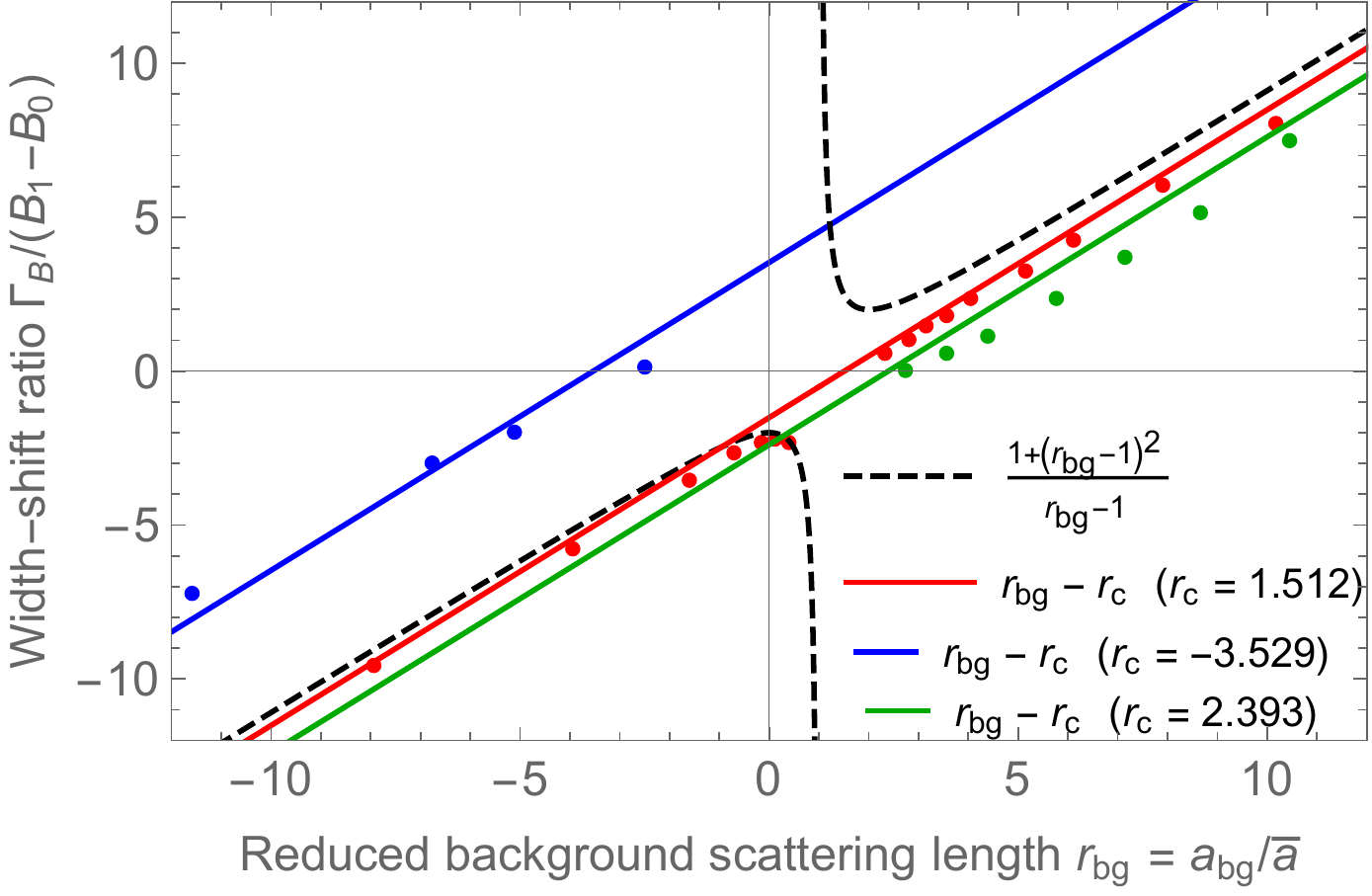}

\caption{\label{fig:Ratio}Ratio between the magnetic width $\Gamma_{B}$ and
shift $B_{1}-B_{0}$, as a function of the reduced background scattering
length $r_{\text{bg}}=a_{\text{bg}}/\bar{a}$, for different values
of the reduced closed-channel scattering length $r_{{\rm c}}=a_{{\rm c}}/\bar{a}$.
The dots are from a multi-channel calculation (see Fig.~\ref{fig:WidthAndPosition}),
and the solid lines correspond to Eq.~(\ref{eq:RatioAdiabatic}),
while the dashed curve corresponds to Eq.~(\ref{eq:PaulFormula}).}
\end{figure}
For fixed values of ${a_{{\rm s}}}$, we can extract and plot the
magnetic width ${\Gamma_{B}}$ as a function of the background scattering
length ${a_{{\rm bg}}}$, which is varied by varying ${a_{{\rm t}}}$.
This is shown in the top panels of Fig.~\ref{fig:WidthAndPosition}.
Using the relation between the magnetic width and the energy width
of the resonance in Eq.~\eqref{eq:magnetic_width-shift}, the data
can be well fitted by the adiabatic formula~(\ref{eq:WidthAdiabatic})
where ${a_{{\rm c}}}$ is set to ${a_{{\rm s}}}$. This indicates
that the resonance is in the adiabatic regime, and that the closed-channel
molecular state is indeed controlled by the singlet scattering length
${a_{{\rm s}}}$, with $a_{{\rm c}}\approx a_{{\rm s}}$. A different
value of ${\overline{\mathcal{W}}}$ has to be set for each value
of ${a_{{\rm s}}}$ indicating that the coupling in the Hamiltonian
is somehow modified when ${a_{{\rm s}}}$ is changed.

Next, for fixed values of ${a_{{\rm s}}}$, we can calculate the last
singlet bound state energy, and find the magnetic field intensity
$B_{1}$ at which the scattering threshold intersects that energy.
In addition, we can extract the resonance position ${B_{0}}$ and
plot it as a function of ${a_{{\rm bg}}}$. This is shown in the lower
panels of Fig.~\ref{fig:WidthAndPosition}. The resonance position
${B_{0}}$ varies approximately linearly with ${a_{{\rm bg}}}$ and
intersects the value $B_{1}$ at $a_{\text{bg}}=a_{\text{c}}$, as
expected from Eq.~(\ref{eq:ShiftAdiabatic}). Moreover, for each
case, the slope of that linear dependence is consistent with the coefficient
of the quadratic dependence of the width parameter, in agreement with
Eqs.~(\ref{eq:WidthAdiabatic}-\ref{eq:ShiftAdiabatic}).

The ratio $\Gamma_{B}/(B_{1}-B_{0})$ of Eq.~(\ref{eq:ratio_Delta-Gamma})
is plotted in Fig.~\ref{fig:Ratio} as a function of the background
scattering length $a_{\text{bg}}$. The linear variation of Eq.~(\ref{eq:RatioAdiabatic})
is again validated and the explicit dependence of the results on the
closed-channel scattering length ${a_{{\rm c}}}$ confirms the inadequacy
of Eq.~(\ref{eq:PaulFormula}), which only depends on the background
scattering length ${a_{{\rm bg}}}$. Although the results presented
here support the validity of Eqs.~(\ref{eq:WidthAdiabatic}-\ref{eq:RatioAdiabatic}),
a full confirmation of these equations in the multi-channel case will
be possible when a reliable way of determining the effective underlying
two-channel model of a resonance is achieved, a task we leave as a
future challenge.\footnote{We note that the determination of an effective two-channel model from
a multi-channel problem was demonstrated in a particular case~\cite{Nygaard2006},
but this determination explicitly made use of the generally incorrect
formula Eq.~(\ref{eq:PaulFormula}) to construct the bare molecular
state of the two-channel model.}

\section{Conclusion}

This work has clarified the relationship between the width and shift
of Fano-Feshbach resonances for van der Waals interactions. This insight
will be crucial for the construction of effective interactions that
can be used to treat few- or many-body problems, while faithfully
reproducing the physics of Fano-Feshbach resonances. Experimentally,
the determination of the shift is possible, as demonstrated recently~\cite{Thomas2018}.
The proposal in Ref.~\cite{Marcelis2008a} for experimentally modifying
the background scattering length of a given resonance is also a promising
direction. This opens interesting perspectives to confirm our result.
A similar analysis of the width and shift could also be of importance
for resonances whose coupling can be controlled, such as microwave
Fano-Feshbach resonances \cite{Papoular2010}.

\noindent %
\noindent\begin{minipage}[t]{1\columnwidth}%
\rule[0.5ex]{1\columnwidth}{1pt}%
\end{minipage}

\subsection*{Acknowledgments}

The authors would like to thank Paul S. Julienne, Eite Tiesinga, Servaas
Kokkelmans, and Maurice Raoult for helpful discussions. P. N. acknowledges
support from the RIKEN Incentive Research Project and JSPS Grants-in-Aid
for Scientific Research on Innovative Areas (No. JP18H05407).

\noindent \hypersetup{urlcolor=myred}{\small{}\bibliographystyle{IEEEtran2}
\bibliography{paper30}
}{\small\par}

\newpage{}

\section*{Appendix 1}

The two-channel Hamiltonian of Eqs.~(\ref{eq:CC1}-\ref{eq:CC2})
is
\begin{equation}
H=\left(\begin{array}{cc}
T & 0\\
0 & T
\end{array}\right)+V\label{eq:HamiltonianDiabatic}
\end{equation}
where $T$ is the kinetic operator $T=-\frac{\hbar^{2}}{2\mu}\frac{d^{2}}{dR^{2}}$
and $V$ is the potential matrix $\left(\begin{array}{cc}
V_{\text{oo}} & V_{\text{oc}}\\
V_{\text{co}} & V_{\text{cc}}
\end{array}\right)$. This expression of the Hamiltonian corresponds to the diabatic basis,
which is a convenient representation for weak coupling $V_{\text{oc}}$.
In the strong-coupling limit, the criterion of Eq.~(\ref{eq:small-coupling})
is not verified and this representation becomes inconvenient. Instead,
we consider the adiabatic basis obtained by diagonalising the potential
matrix,
\begin{equation}
V^{\prime}=P^{-1}VP=\left(\begin{array}{cc}
V_{+} & 0\\
0 & V_{-}
\end{array}\right)\label{eq:AdiabaticPotentialMatrix}
\end{equation}
where 
\begin{equation}
V_{\pm}=\frac{V_{\text{oo}}+V_{\text{cc}}}{2}\pm\sqrt{\frac{(V_{\text{oo}}-V_{\text{cc}})^{2}}{4}+\vert V_{\text{oc}}\vert^{2}}\label{eq:AdiabaticPotentials}
\end{equation}
Assuming that $V$ is real, the transformation matrix $P$ is given
by
\begin{equation}
P=\left(\begin{array}{cc}
\cos\frac{\theta}{2} & -\sin\frac{\theta}{2}\\
\sin\frac{\theta}{2} & \cos\frac{\theta}{2}
\end{array}\right)\qquad\text{with}\quad\tan\theta=\frac{2V_{\text{oc}}}{V_{\text{oo}}-V_{\text{cc}}}\label{eq:TransformationMatrix}
\end{equation}
where $0<\theta<\pi$. Although the potential becomes diagonal in
the adiabatic basis, the kinetic operator transforms as follows:
\begin{equation}
P^{-1}TP=\left(\begin{array}{cc}
T+\frac{\hbar^{2}}{2\mu}Q^{2} & -\frac{\hbar^{2}}{2\mu}\left[\frac{dQ}{dR}+2Q\frac{d}{dR}\right]\\
\frac{\hbar^{2}}{2\mu}\left[\frac{dQ}{dR}+2Q\frac{d}{dR}\right] & T+\frac{\hbar^{2}}{2\mu}Q^{2}
\end{array}\right)\label{eq:HamiltonianAdiabatic}
\end{equation}
where the terms $Q(R)=-\frac{1}{2}\frac{d\theta(R)}{dR}$ arise from
the action of the derivative operator in $T$. This gives rise to
new off-diagonal couplings called radial couplings.

For magnetic Fano-Feshbach resonances of alkali atoms, the function
$Q(R)$ shows a peak that can be located in the van der Waals region~\cite{Mies1996}.

{\small{}}{\small\par}

\section*{Appendix 2}

\paragraph{Calculation of ${Y_{ij}}$ \textendash{}}

From Eqs.~(\ref{eq:CC1}-\ref{eq:CC2}) we have 
\begin{align*}
\psi_{{\rm o}}(R) & =A\times\hat{f}_{{\rm o}}(R)+\int_{0}^{\infty}dR^{\prime}{\hat{\mathcal{G}}}_{{\rm o}}^{E}(R,R^{\prime})V_{{\rm oc}}(R^{\prime})\psi_{{\rm c}}(R^{\prime})\\
\psi_{{\rm c}}(R) & =B\times\hat{f}_{{\rm c}}(R)+\int_{0}^{\infty}dR^{\prime}{\hat{\mathcal{G}}}_{{\rm c}}^{E}(R,R^{\prime})V_{{\rm co}}(R^{\prime})\psi_{{\rm o}}(R^{\prime}),
\end{align*}
where ${A}$ and ${B}$ are two numbers, and we have introduced the
two Green's function, 
\begin{equation}
\hat{\mathcal{G}}_{i}^{E}(R,R^{\prime})=\frac{2\mu}{\hbar^{2}}\frac{1}{W[\hat{f}_{i},\hat{g}_{i}]}\begin{cases}
\hat{f}_{i}(R)\hat{g}_{i}(R^{\prime}) & \text{for }R<R^{\prime}\\
\hat{f}_{i}(R^{\prime})\hat{g}_{i}(R) & \text{for }R>R^{\prime}
\end{cases}
\end{equation}
satisfying the radial equation, 
\begin{equation}
\left(-\frac{\hbar^{2}}{2\mu}\frac{d^{2}}{dR^{2}}+V_{ii}(R)-E\right){\hat{\mathcal{G}}}_{i}^{E}(R,R^{\prime})=-\delta(R-R^{\prime})
\end{equation}
and the appropriate boundary condition ${{\mathcal{G}}_{i}^{E}(R,R^{\prime})\xrightarrow[R\to0]{}0}$,
since both ${\hat{f}_{i}}$ are regular at the origin. This gives
\begin{align*}
\psi_{{\rm o}}(R) & =A\times\hat{f}_{{\rm o}}(R)+(\hat{f}_{{\rm o}}\vert\psi_{{\rm c}})_{<}\hat{g}_{{\rm o}}(R)+(g_{{\rm o}}\vert\psi_{{\rm c}})_{>}\hat{f}_{{\rm o}}(R)\\
\psi_{{\rm c}}(R) & =B\times\hat{f}_{{\rm c}}(R)+(\hat{f}_{{\rm c}}\vert\psi_{{\rm o}})_{<}\hat{g}_{{\rm c}}(R)+(g_{{\rm c}}\vert\psi_{{\rm o}})_{>}\hat{f}_{{\rm c}}(R).
\end{align*}
Therefore, for ${R>R_{{\rm free}}}$, 
\begin{align*}
\psi_{{\rm o}}(R) & \xrightarrow[R\gtrsim R_{{\rm free}}]{}A\times\hat{f}_{{\rm o}}(R)+(\hat{f}_{{\rm o}}\vert\psi_{{\rm c}})\hat{g}_{{\rm o}}(R)\\
\psi_{{\rm c}}(R) & \xrightarrow[R\gtrsim R_{{\rm free}}]{}B\times\hat{f}_{{\rm c}}(R)+(\hat{f}_{{\rm c}}\vert\psi_{{\rm o}})\hat{g}_{{\rm c}}(R).
\end{align*}
We find the two linearly independent solutions ${\psi^{(1)}}$ and
${\psi^{(2)}}$ for ${(A,B)=(1,0)}$ and ${(A,B)=(0,1)}$. For ${(A,B)=(1,0)}$,
we get 
\begin{equation}
Y_{{\rm oo}}=-(\hat{f}_{{\rm o}}\vert\psi_{{\rm c}}^{(1)})\quad;\quad Y_{{\rm co}}=-(\hat{f}_{{\rm c}}\vert\psi_{{\rm o}}^{(1)}),
\end{equation}
and for ${(A,B)=(0,1)}$, we get 
\begin{equation}
Y_{{\rm oc}}=-(\hat{f}_{{\rm o}}\vert\psi_{{\rm c}}^{(2)})\quad;\quad Y_{{\rm cc}}=-(\hat{f}_{{\rm c}}\vert\psi_{{\rm o}}^{(2)}).
\end{equation}

\paragraph{Limit of weak coupling}

In the limit of weak coupling, one can make the following approximations:
\begin{align}
Y_{{\rm co}} & =-(\hat{f}_{{\rm c}}\vert\psi_{{\rm o}}^{(1)})\approx-(\hat{f}_{{\rm c}}\vert\hat{f}_{{\rm o}})\\
\psi_{{\rm c}}^{(1)} & =\hat{g}_{{\rm c}}(\hat{f}_{{\rm c}}\vert\psi_{{\rm o}}^{(1)})_{<}+\hat{f}_{{\rm c}}(g_{{\rm c}}\vert\psi_{{\rm o}}^{(1)})_{>}\\
 & \approx\hat{g}_{{\rm c}}(\hat{f}_{{\rm c}}\vert\hat{f}_{{\rm o}})_{<}+\hat{f}_{{\rm c}}(g_{{\rm c}}\vert\hat{f}_{{\rm o}})_{>}\nonumber 
\end{align}
and therefore,
\begin{equation}
Y_{{\rm oo}}=-(\hat{f}_{{\rm o}}\vert\psi_{{\rm c}}^{(1)})=-\Big(\hat{f}_{{\rm o}}\Big\vert\hat{g}_{{\rm c}}(\hat{f}_{{\rm c}}\vert\hat{f}_{{\rm o}})_{<}+\hat{f}_{{\rm c}}(g_{{\rm c}}\vert\hat{f}_{{\rm o}})_{>}\Big).
\end{equation}
Likewise, 
\begin{align}
Y_{{\rm oc}} & =-(\hat{f}_{{\rm o}}\vert\psi_{{\rm c}}^{(2)})\approx-(\hat{f}_{{\rm o}}\vert\hat{f}_{{\rm c}})\\
\psi_{{\rm o}}^{(2)} & =\hat{g}_{{\rm o}}(\hat{f}_{{\rm o}}\vert\psi_{{\rm c}})_{<}+\hat{f}_{{\rm o}}(g_{{\rm o}}\vert\psi_{{\rm c}})_{>}\\
 & \approx\hat{g}_{{\rm o}}(\hat{f}_{{\rm o}}\vert\hat{f}_{{\rm c}})_{<}+\hat{f}_{{\rm o}}(g_{{\rm o}}\vert\hat{f}_{{\rm c}})_{>}\nonumber 
\end{align}
and therefore,
\begin{equation}
Y_{{\rm cc}}=-(\hat{f}_{{\rm c}}\vert\psi_{{\rm o}}^{(2)})\approx-\Big(\hat{f}_{{\rm c}}\Big\vert\hat{g}_{{\rm o}}(\hat{f}_{{\rm o}}\vert\hat{f}_{{\rm c}})_{<}+\hat{f}_{{\rm o}}(g_{{\rm o}}\vert\hat{f}_{{\rm c}})_{>}\Big).
\end{equation}

\section*{Appendix 3}

In this appendix, we establish the connection between the coefficient
$\alpha$ of Eq.~\eqref{eq:TaylorExpansion} and the normalisation
of the bound-state wave function $\hat{f}_{m}$ of section \ref{sec:6}.
The exponentially convergent function ${\psi_{{\rm c}}\approx\hat{f}_{{\rm c}}+\alpha(E-E_{m})\hat{g}_{{\rm c}}}$
becomes the closed-channel bound state ${\hat{f}_{m}}$ when $E=E_{m}$.
The functions ${\psi_{{\rm c}}}$ and ${\hat{f}_{m}}$ are solutions
of the closed-channel radial equations: 
\begin{align}
 & -\frac{\hbar^{2}}{2\mu}\psi_{{\rm c}}^{\prime\prime}+(V_{{\rm cc}}-E)\psi_{{\rm c}}=0\\
 & -\frac{\hbar^{2}}{2\mu}\hat{f}_{m}^{\prime\prime}+(V_{{\rm cc}}-E_{m})\hat{f}_{m}=0.
\end{align}
Multiplying the first equation by ${\hat{f}_{m}}$ and the second
equation by ${\psi_{{\rm c}}}$, taking the difference between the
two equations, and integrating gives
\begin{multline}
-\frac{\hbar^{2}}{2\mu}\int_{0}^{\infty}\left(\psi_{{\rm c}}^{\prime\prime}\hat{f}_{m}-\hat{f}_{m}^{\prime\prime}\psi_{{\rm c}}\right)dR\\
-(E-E_{m})\int_{0}^{\infty}\psi_{{\rm c}}\hat{f}_{m}dR=0.
\end{multline}
Integrating by parts gives 
\begin{equation}
\frac{\hbar^{2}}{2\mu}\left[\psi_{{\rm c}}^{\prime}\hat{f}_{m}-\hat{f}_{m}^{\prime}\psi_{{\rm c}}\right]_{R=0}-(E-E_{m})\int_{0}^{\infty}\psi_{{\rm c}}\hat{f}_{m}dR=0.
\end{equation}
Using the explicit form of ${\psi_{{\rm c}}}$, 
\begin{align}
\frac{\hbar^{2}}{2\mu}\left[(\hat{f}_{{\rm c}}^{\prime}+\alpha(E-E_{m})\hat{g}_{{\rm c}}^{\prime})\hat{f}_{m}-\hat{f}_{m}^{\prime}(\hat{f}_{{\rm c}}+\alpha(E-E_{m})\hat{g}_{{\rm c}})\right]_{R=0}\nonumber \\
-(E-E_{m})\int_{0}^{\infty}(\hat{f}_{{\rm c}}+\alpha(E-E_{m})\hat{g}_{{\rm c}})\hat{f}_{m}dR=0.
\end{align}
Using the fact that ${\hat{f}_{m}(0)=0}$ and ${\hat{f}_{{\rm c}}(0)=0}$,
we obtain 
\begin{equation}
\int_{0}^{\infty}(\hat{f}_{{\rm c}}+\alpha(E-E_{m})\hat{g}_{{\rm c}})\hat{f}_{m}dR=\frac{\hbar^{2}}{2\mu}\alpha W[\hat{f}_{m},\hat{g}_{{\rm c}}].
\end{equation}
Finally, taking the limit ${E\to E_{m}}$, we get ${\hat{f}_{{\rm c}}\to\hat{f}_{m}}$,
and ${W[\hat{f}_{m},\hat{g}_{{\rm c}}]\to W[\hat{f}_{m},\hat{g}_{m}]=1}$,
which gives
\begin{equation}
\int_{0}^{\infty}\hat{f}_{m}^{2}dR=\frac{\hbar^{2}}{2\mu}\alpha.
\end{equation}
This last equation relates the coefficient ${\alpha}$ to the normalisation
of the bound state wave function ${\hat{f}_{m}}$. Thus, for ${E\approx E_{m}}$,
the state ${\Psi_{{\rm c}}(R)=\frac{\hat{f}_{{\rm c}}(R)}{R}\sqrt{\frac{2\mu}{4\pi\hbar^{2}\alpha}}}$
is approximately the bound state ${\Psi_{m}(R)=\frac{\hat{f}_{m}(R)}{R}\sqrt{\frac{2\mu}{4\pi\hbar^{2}\alpha}}}$,
with the proper normalisation ${\int_{0}^{\infty}4\pi R^{2}dR\vert\Psi_{m}(R)\vert^{2}=1}$.
\end{document}